\documentclass[pdflatex]{sn-jnl}

\usepackage{algorithm}
\usepackage{algorithmic}
\usepackage{graphicx}%
\usepackage{multirow}%
\usepackage{amsmath,amssymb,amsfonts}
\usepackage{amsthm}%
\usepackage{mathrsfs}%
\usepackage[title]{appendix}%
\usepackage{xcolor}%
\usepackage{hyperref}
\usepackage{textcomp}%
\usepackage{manyfoot}%
\usepackage{booktabs}%
\usepackage{algorithm}%
\usepackage{amssymb}
\usepackage{listings}%
\usepackage{subfigure}%
\usepackage{caption}%
\usepackage{lipsum}%
\usepackage{pgfplotstable}
\usepackage{tikz}%
\usepackage{tkz-tab}%
\usepackage{latexsym}%
\usepackage{pgfplots}%
\usepackage[utf8]{inputenc}%
\usetikzlibrary{patterns}%
\usepgfplotslibrary{statistics}%
\pgfmathdeclarefunction{fpumod}{2}{%
    \pgfmathfloatdivide{#1}{#2}%
    \pgfmathfloatint{\pgfmathresult}%
    \pgfmathfloatmultiply{\pgfmathresult}{#2}%
    \pgfmathfloatsubtract{#1}{\pgfmathresult}%
    \pgfmathfloatifapproxequalrel{\pgfmathresult}{#2}{\def\pgfmathresult{5}}{}%
}
\pgfplotsset{boxplot legend/.style={
    legend image code/.code={
        \draw[#1] (0cm,-0.1cm) rectangle (0.4cm,0.1cm)
        (0.2cm,-0.1cm) -- (0.2cm,-0.2cm) (0.05cm,-0.2cm) -- (0.35cm,-0.2cm)
        (0.2cm,0.1cm) -- (0.2cm,0.2cm) (0.05cm,0.2cm) -- (0.35cm,0.2cm);
     \path (0cm,0.24cm) (0cm,-0.24cm);  
    },
}}

\pgfplotsset{width=9cm,compat=1.5.1}

\begin{document}

\title[Article Title]{MAIDS: Malicious Agent Identification-based Data Security Model for Cloud Environments}


\author*[1]{\fnm{Kishu} \sur{Gupta}}\email{kishuguptares@gmail.com}
\equalcont{These authors contributed equally to this work.}

\author[2,3]{\fnm{Deepika} \sur{Saxena}}\email{deepika@u-aizu.ac.jp}
\equalcont{These authors contributed equally to this work.}

\author[4]{\fnm{Rishabh} \sur{Gupta}}\email{rishabhgpt66@gmail.com}
\equalcont{These authors contributed equally to this work.}

\author[3,5]{\fnm{Ashutosh Kumar} \sur{Singh}}\email{ashutosh@iiitbhopal.ac.in}
\equalcont{These authors contributed equally to this work.}

\affil*[1]{\orgdiv{Department of Computer Science \& Engineering}, \orgname{National Sun Yat-sen University}, \orgaddress{\city{Kaohsiung}, \postcode{80424}, \country{Taiwan}}}
\affil[2]{\orgdiv{Department of Computer Science \& Engineering}, \orgname{University of Aizu}, \orgaddress{\city{Aizuwakamatsu}, \state{Fukushima}, \country{Japan}}}
\affil[3]{\orgdiv{Department of Computer Science},\orgname{The University of Economics and Human Sciences}, \orgaddress{\city{Warsaw}, \postcode{01043}, \country{Poland}}}
\affil[4]{\orgdiv{Department of Computer Applications},\orgname{SRM Institute of Science \& Technology, Delhi-NCR Campus, Modinagar}, \orgaddress{\city{Ghaziabad}, \postcode{201204}, \state{Uttar Pradesh}, \country{India}}}
\affil[5]{\orgdiv{Department of Computer Science \& Engineering}, \orgname{Indian Institute of Information Technology}, \orgaddress{\city{Bhopal}, \postcode{462003}, \state{Madhya Pradesh}, \country{India}}}


\abstract{With the vigorous development of cloud computing, most organizations have shifted their data and applications to the cloud environment for storage, computation, and sharing purposes. During storage and data sharing across the participating entities, a malicious agent may gain access to outsourced data from the cloud environment. A malicious agent is an entity that deliberately breaches the data. This information accessed might be misused or revealed to unauthorized parties. Therefore, data protection and prediction of malicious agents have become a demanding task that needs to be addressed appropriately. To deal with this crucial and challenging issue, this paper presents a \textbf{M}alicious \textbf{A}gent \textbf{I}dentification-based \textbf{D}ata \textbf{S}ecurity (\textbf{MAIDS}) Model which utilizes XGBoost machine learning classification algorithm for securing data allocation and communication among different participating entities in the cloud system. The proposed model explores and computes intended multiple security parameters associated with online data communication or transactions. Correspondingly, a security-focused knowledge database is produced for developing the XGBoost Classifier-based Malicious Agent Prediction (XC-MAP) unit. Unlike the existing approaches, which only identify malicious agents after data leaks, MAIDS proactively identifies malicious agents by examining their eligibility for respective data access. In this way, the model provides a comprehensive solution to safeguard crucial data from both intentional and non-intentional breaches, by granting data to authorized agents only by evaluating the agent's behavior and predicting the malicious agent before granting data. 

The performance of the proposed model is thoroughly evaluated by accomplishing extensive experiments, and the results signify that the MAIDS model predicts the malicious agents with high \textit{accuracy}, \textit{precision}, \textit{recall}, and \textit{F1-scores} up to 95.55\%, 95.30\%, 95.50\%, and 95.20\%, respectively. This enormously enhances the system's security in terms of authorized data access accuracy up to 55.49\%, precision up to 43.15\%, recall up to 55.49\%, and f1-score up to 39.96\%, respectively, as compared to state-of-the-art work.
}



\maketitle
\begin{center}
	\footnotesize{\textbf{This article has been accepted for publication in Cluster Computing Journal © under exclusive licence to Springer Science+Business Media, LLC, part of Springer Nature 20242024. \\Citation information: DOI 10.1007/s10586-023-04263-9.\\ Received: 25 September 2023 / Revised: 10 December 2023 / Accepted: 28 December 2023 / Published online: 23 February 2024. \\Personal use of this material is permitted. Permission from Springer must be obtained for all other uses, in any current or future media, including reprinting/republishing this material for advertising or promotional purposes, creating new collective works, for resale or redistribution to servers or lists, or reuse of any copyrighted component of this work in other works. This work is freely available for survey and citation.}}
\end{center}
\section{Introduction}\label{sec1}
Cloud-assisted data sharing has appeared as an essential service for any organization to enhance the utility of the shared data \cite{song2022public, saxena2022high}. Due to the considerable benefits of the cloud, including tremendous computation capacity, ease-of-access and massive storage space at an reasonable cost \cite{wei2016secure, singh2021quantum}, most organizations are pushing a substantial amount of data from on-premise to cloud platforms for distinct pursuits such as huge storage, massive analysis, and further sharing with multiple parties and stakeholders for utilization \cite{gupta2022iot, saxena2021secure, icdam2021}. Currently, 94\% of organizations use cloud services, according to the Cloud Adoption Statistics report \cite{cloud}. However, relying on the third-party's cloud platform is not prudent, especially for confidential, sensitive, and crucial data, because of the reason that the proprietors lose control of the data outsourced to the cloud \cite{gupta2022differential, chandra2021, FedMUP}. Consequently, on one side data proprietors might be reluctant to upload their data to the cloud due to attractive storage and analysis facilities \cite{yin2022efficient, rgupta_singh2022}. However, on the other side, the cloud itself could be manipulated and exploited to reveal outsourced data to third parties for illicit purposes \cite{shen2018enabling, li2019meta}. In 2022, the average size and cost of data breaches grew by 2.6\% and 12.7\%, respectively, over the previous year, and the likelihood of having breached data will rise to 14.94\% by 2023 \cite{IBMPonemon}. In this way, data compromises have emerged as a primary obstacle for the organization responsible for managing it. A prominent way to resolve this impending paramount problem is to identify the malicious agent responsible for the data leak.  

To catch malicious agents \cite{martinsurvey2021, reviewer}, several strategies have been proposed that can be extensively classified into two broad categories (i) watermarking and (ii) probability-based approaches. The watermarking approach ingrains a unique code into each shared copy of the data. The embedded code is retrieved from the acknowledged leaked document to identify the culprit. Although, it is an adequate method for identifying malicious agents, the content that has been watermarked is susceptible to change \cite{shehab2007watermarking, saxena2021osc, almehmadi2022novel}. Employing a probabilistic approach makes it possible to predict the likelihood that a single agent, a group of agents, or an illegal entity has exposed the data or obtained the data directly from an individual agent. This approach may be quite practical in some situations, but it cannot locate the real thing \cite{gupta2020mlpam, rgupta2022, ML2Woznica2023, ML1Huang2022}. Moreover, the existing approaches identify the malicious agent after the occurrence of a data leak, which is inadequate in the real environment. To the best of the authors’ knowledge, no existing models predict malicious agents before data breaches along with secure data storage and sharing.

To tackle the challenge mentioned earlier, a novel model named a "\textbf{M}alicious \textbf{A}gent \textbf{I}dentification-based \textbf{D}ata \textbf{S}ecurity Model (\textbf{MAIDS})" is proposed that examines the request of the agents as ‘malicious’ or ‘non-malicious’ for protecting data distribution and transmission for secure communication. The presented model improvises two units named \textit{Agent Eligibility Estimation} (AEE) and \textit{XGBoost Machine Learning-based Malicious Agent Prediction} (XC-MAP). Agent eligibility is estimated by considering a variety of pertinent security parameters related to each data request and the corresponding computed scores. The data values produced with an AEE unit are then utilized to train an XC-MAP unit. Accordingly, by providing predictions of data request intentions and apparent harmful activities before the occurrence, the MAIDS model aids in strengthening data distribution and communication security in the cloud environment.
\begin{figure}[!htbp]%
\centering
\includegraphics[width=0.9\textwidth]{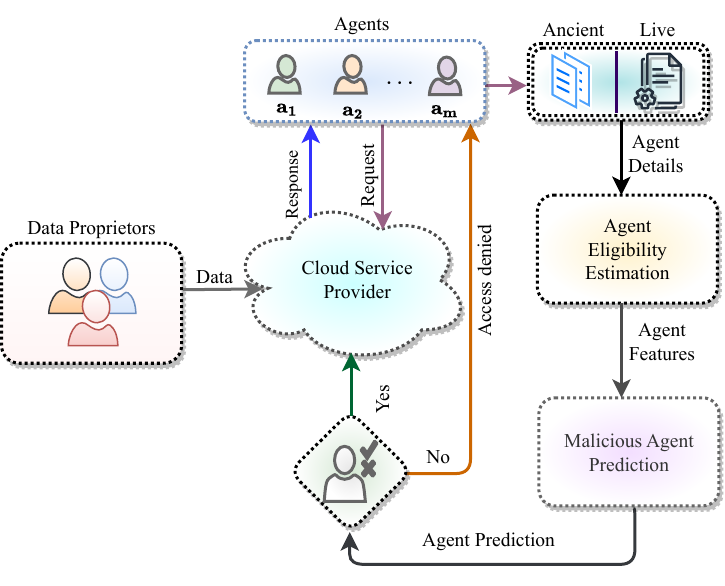}
\caption{Operational flow of proposed MAIDS Model}\label{op_flow}
\end{figure}

A detailed operational flow of the presented model is demonstrated in Fig. \ref{op_flow}, where data proprietors generate the data and store it in the cloud for computing, analysis, sharing, and distribution among relevant stakeholders. For the sake of data security and to prohibit the occurrence of data breaches by any malicious agent, agent eligibility is computed for examining the respective data request as 'malicious' or 'non-malicious' before the assignment of data access permission during allocation and communication. The key contributions of MAIDS are highlighted as follows:
\begin{itemize}
\item[$\bullet$] A novel concept known as \textit{Agent Eligibility Estimation} (AEE) is devised to determine the potential intentions of agents seeking cloud data for diverse pursuits. Each data request's security score is computed and produced through an AEE unit by considering several \textit{critical security parameters}.
\item[$\bullet$] \textit{A Malicious Agent Prediction} Unit based on XGBoost Classification (XC-MAP) is designed for proactive real-time identification of malicious agents and data demands.
\item[$\bullet$] MAIDS facilitates various data proprietors to share
outsourced data readily. Each agent's behavior is anticipated to prevent data breaches.
\item[$\bullet$] The experiments have been carried out utilizing a variety of data sets, and feature analysis is accomplished to assess the effectiveness of the proposed model. The acquired outcomes are compared with state-of-the-art works in terms of accuracy, precision, recall, and F1 score.
\end{itemize}
\par Table \ref{TableTerminology} showcases a description of the terminologies employed throughout this article.
\begin{table}[!htbp]
\caption{Snapshot of Terminologies employed in the MAIDS model}
\label{TableTerminology}
\centering
\begin{tabular}{l l|l l}\hline \hline
\textbf{Term:} &  \textbf{Definition} &  \textbf{Term:} &  \textbf{Definition} \\ \hline \hline

\textit{${DP}$}: &\texttt{Data Proprietor} & $n$: &\texttt{Number of Data Proprietor} \\

\textit{$\theta$}: & \texttt{Data Leakage Status} & \textit{NE}: & \texttt{Non-trusted Entity} \\

\textit{$t$}: & \texttt{Time} & \textit{${SP}$}: & \texttt{Security Risk Parameter} \\ 

\textit{$a^{hd}$}: & \texttt{Historical Data Access Information} & \textit{$\psi^{a}$}: & \texttt{Agent Live Details}  \\ 

\textit{$PR$}: & \texttt{Predicted Accuracy} & \textit{$PP$}: & \texttt{Predicted Precision} \\

\textit{$PR$}: & \texttt{Predicted Recall Data} & \textit{$PF$}: & \texttt{Predicted F1-Score}  \\

\textit{${L}^{Int}$}: & \texttt{Unauthorised Data Leakages} & \textit{${ACD}$}: & \texttt{Attack Component Details} \\

\textit{$\Check{Thr}^{freq}$}: & \texttt{Threshold Frequency} & \textit{$w$}: & \texttt{Weight ($leaf$ score)} \\

\textit{${DO}$}: & \texttt{Data Objects} & \textit{$n^\ast$}: & \texttt{Number of Data Objects} \\

\textit{${A}$}: & \texttt{Agents} & \textit{$m$}: & \texttt{Number of Agents} \\

\textit{$AF$}: & \texttt{Agent Features} & \textit{${DA}^{gross}$}: & \texttt{Total Data Access} \\   

\textit{$N$}: & \texttt{Total Security Parameters}  & \textit{$\wp$}: & \texttt{Security Risk Information} \\

\textit{${PD}$}: & \texttt{Prohibited Data} & \textit{$do_{j}$}: & \texttt{Requesting Data Objects} \\

\textit{${PDO}$}: & \texttt{Permissible Data Objects} & \textit{$z$}: & \texttt{No. of Requesting Data} \\

\textit{$\lambda$}: & \texttt{Shrinkage parameter} & \textit{$\gamma$}: & \texttt{Minimum loss reduction Gain} \\

\textit{$\zeta$}: & \texttt{Loss Function value} & \textit{$T$}: & \texttt{Total Leaf Nodes in a Tree} \\

\textit{$q$}: & \texttt{Tree Structure} & \textit{$l$}: & \texttt{Differential loss function} \\

\textit{$\Omega$}: & \texttt{Penalisation Factor} & \textit{$\hat{y}^t_i$}: & \texttt{Predicted value} \\

\textit{$g_i$}: & \texttt{First Order Derivative} & \textit{$h_i$}: & \texttt{Second Order Derivative} \\
\hline \hline
\end{tabular}
\end{table}

The rest of the paper is organized as follows:  Section \ref{sec2} comprises a brief review of methodologies that have been proposed for malicious entity prediction together with a comparison table reflecting existing work done so far. Section \ref{sec3} discusses the XGBoost-driven proposed MAIDS model for malicious agent identification to ensure data security. Two different units (i) to analyze the agent behavior by employing agent behavior modeling, and (ii) to predict the malicious agent by taking advantage of the computational goodness of the XGBoost algorithm are explained in Section \ref{sec4} and Section \ref{sec5}, respectively. Furthermore, the operational design and complexity computation for the proposed data security model under consideration is showcased in Section \ref{sec6}. In this line, Section \ref{sec7} elaborates on the experimental work performed for the proposed model including a discussion of the experimental setup, implementation, and dataset bringing forward the corresponding results and comparison with state-of-the-art methods. Ultimately, Section \ref{sec8} presents the conclusion drawn from the proposed MAIDS model.
\section{Related Work}\label{sec2}
To ensure data security abundant models are studied and defined. Broadly data security can be imparted in a reactive manner i.e. once after the occurrence of a data breach event on the other hand persuade security task proactively to ensure identification of possible leakers even before the occurrence of a data breach event through agent behaviour analysis.

Papadimitriou et al. \cite{5487521} proposed a guilty agent detection (GAM) model. This model tries to unveil the suspicious agent who seems to have leaked the extremely sensitive content. Model employs fake data object addition into original data techniques like watermarking and provenance, to highlight the possible culprit. To analyze what seems to be guilty, a statistical approach has been used. The major drawback this model suffers from is that it works for explicit type of data request only i.e. the scenario in which agent demand for particular data elements.

Rohde et al. \cite{Rohde2018} presented a model for early-stage dynamic malware detection. This model tries to determine whether an executable payload is malicious using ensemble recurrent neural networks (RNNs) over behavioral data. The model is the first one to determine the malicious executable proactively during execution within 5 seconds, rather than post-execution. The most crucial drawback the model faces is that it does not showcase an integrated approach to predict file-specific behavioral detection across dissimilar operating systems.

Sharif et al. \cite{Sharif2018} proposed a system for proactive identification of malicious content over the web. The model employed machine learning classification tools to analyze web browsing for the HTTP data generated in 3 months duration for over twenty thousand users. The model can predict malicious exposure even before it occurs just by observing long-term behavior in a single browsing session for self-reported features, contextual features, and, past behavior features. Though the model successfully predicts malicious content proactively but lacks accuracy for behavioural data. 

Gupta et al. \cite{gupta2019dynamic} came up with data leaker 
identification model based on a dynamic threshold (DT-ILIS). This model tries to determine the data leaker using various access control mechanisms. Moreover, data distribution pattern among various agents is also taken care of by this model. Though this model showcases improved parameters like high probability, success rate, and detection rate, the model assumes the distributor is a trusted party which seems practically not possible in all cases.

A data security model to find out the social bots and most influential users on social networking sites is proposed by \cite{Lingam2019}. To detect the social bots author employed Deep Q Learning (DQL) based on several social features like tweet-related features, user profile-related features, and, social graph-related features. Though this model successfully identifies the social bots and find the most influential user as well, however, it lags in term of real-time application scenario.

A dynamic approach to identify online information leakers (On-ILIS) is introduced in \cite{guptaOnILIS}. This model explains a very prominent approach to distributing the data to satisfy the agent(s) requirements. The approach provides good improvement for data disclosure risk but it suffers from a lack of data privacy.

Rabbani et al. \cite{RABBANI2020102507} devised a new approach to raise cloud service provider potential to analyze the user's behavior for data security. This model employs particle swarm optimization probabilistic neural network (PSO-PNN); a self-optimised machine learning approach to learn the activity pattern of users. This collective approach successfully and efficiently classifies the user activity as normal or suspicious. However, the approach still lacks raw data handling with heavy network traffic. 

Gupta et al. \cite{gupta2020mlpam} presented a machine learning-oriented malicious user identification model using a probabilistic approach (ML-PAM). This model deploys some probabilistic strategies to find malicious users considering the data distribution pattern among them. Moreover, the model ensured data protection and privacy by way of cipher text policy attribute-oriented encryption and differential privacy mechanisms. However, one of the improvements of this model can be sought in terms of proactive measures as this model look out for guilty one after the occurrence of leakage.

A learning-oriented data leakage prevention model using the XGBoost classification technique is devised by \cite{gupta_Kush_2020}. This model proposed an ensemble XGBoost version of existing XGBoost classification techniques. Though the model performance parameters like high accuracy, precision, recall, etc. are remarkable though model works reactively i.e. starts investigation after the happening of event.

Afshar et al. \cite{afshar2021incorporating} give a concept of an access control scheme based on attribute/behavior (ABBAC) to find out the suspicious users. The model grants data access to users based on their behavior. Though the model performs well enough to protect from unauthorized access, however, the model computes very little accuracy that is with limited data access.

Khan et al. \cite{Khan2021} introduced a robust privacy-conserving intrusion detection system (PC-IDS). This model employs two units, the first to pre-process the data employing data cleaning, data reduction, etc, and the other unit to identify the malicious intrusion using a particle swarm optimization-based neural network, respectively. The model exhibits a great enhancement in the detection rate parameters compared with existing works but computational complexity stands high.

Raja et al. \cite{Raja2022} come up with a privacy detection mechanism to identify fake accounts on social network platforms with the intent of data protection. This model utilizes data mining and the SVM-NN approach to identify the fake accounts. Data mining deploys a 3PS (Publically Privacy Protected System) based approach and considers no of shared posts, recent activities, and behavioral patterns to identify malicious accounts. SVM-NN classifies the account as fake or real one. The major impedance this model suffers is the lack of user recommendation preferences into consideration. 

Ranjan et al. \cite{Ranjana2022} highlighted a novel approach for malicious user prediction over web application traffic. Agent behavior analysis is performed for authentication by employing big data analytics to the application layer logs. The model utilizes one of the machine learning classification approaches, the Random Forest Application Algorithm (RFAA) to obtain a prediction. The model successfully identifies the malicious one with 65-70 \% accuracy in real-time scenarios. So critically there is a huge scope for model accuracy and lower the delay time.

Gupta et al. \cite{9865138} proposed a proactive model for malicious user prediction (MUP) before allocating the data to the users. This model employs quantum machine learning (QML) using qubits using the Pauli gate in a multi-layer environment to find out who seems to be a guilty user, in a proactive manner. This is the first article using QML for MUP proactively with better accuracy. But the crucial issue is the model's computational complexity, which makes it harder.

A comparative summary of existing malicious agent prediction (MAP) approaches is highlighted in Table \ref{tabrelworkside}. One common observation recorded from the above-stated models is that they work either in reactive i.e. find out the guilty agent once after happening of leakage event or proactive manner. Moreover, they restrict data access, and data availability, also add some fake data/noise to ensure data security. Data sharing is the key to excel. The need of the hour is to make data sharing available with utmost access and protection but not at the cost of data availability.
\begin{sidewaystable}
\caption{Comparative summary of Malicious Agent Prediction Approaches}\label{tabrelworkside}
\begin{tabular*}{\textheight}{@{\extracolsep\fill}lcccccc}
\toprule%
Contributor & Model/  & Strategy & Dataset & Implementation & Predicted & Result \\
(Timeline) & Approach &&&  & parameters & \\
\midrule \hline
Papadimitriou et & Data allocation  & Agent selection  & --- & Python  & Probability & Guilty \\
al. \cite{5487521} (2011) & using fake objects & and object selection &  & & & user \\ \hline
Rhode et al. \cite{Rohde2018} & RNN & Feed forward & Virus   & Python  & Accuracy  & Malicious \\
(2018) &  & network & Total & 2.7 & & malware \\ \hline
Sharif et al. \cite{Sharif2018} & Regression/ & Feed Forward & RSeBIS   & Python  & True/False  & Malicious \\
(2018) & RF/DNN  & network &  & & Positive Rate & payload \\ \hline
Gupta et al. \cite{gupta2019dynamic} & Threshold based  & Data allocation  & --- & C++  & Average probability, & Malicious entity \\
(2019) & guilt detection & using round robin & & & success rate & (75$\%$ improvement) \\ \hline
Lingam et al. \cite{Lingam2019} & Deep Q Learning  & State-Action  & Twitter & Python   & Accuracy & Social bot, \\
(2019) & (DQL)  & pairs & network & & & influential user \\ \hline 
Gupta et al. \cite{guptaOnILIS} & Date request & Data distribution & UCI & Python & Average & Malicious user \\
(2021) & in online manner & strategy & repository & &probability & (41$\%$ improvement) \\ \hline
Rabbani et al. \cite{RABBANI2020102507} & PSO-PNN & Multilayer feed  & UNSW- & Python  & Precision  & Malicious\\
(2020) &  & forward NN & NB15 & & F measures & user \\ \hline
Gupta et al. \cite{gupta2020mlpam} & Data privacy,  & Encryption, & UCI & Python,  & Detection & Malicious user \\
(2020) & protection & Probabilistic & repository & C++ & accuracy & (186$\%$ improvement) \\ \hline
Gupta et al. \cite{gupta_Kush_2020} & XGBoost   & ensamble  & UCI & Python  & Accuracy  & Malicious\\
(2020) & classifier & learning & repository & & & user \\ \hline
Afshar et al. \cite{afshar2021incorporating} & Attribute based & Analysing real-time & UCI & Python  & RMSE  & Malicious \\
(2021) & access control & $\&$ historical data & repository & & MRMSE & insider user \\ \hline
Khan et al. \cite{Khan2021} & Smart power  & Particle swarm  & UNSW- & Python  & Detection & Malicious \\
(2021) & system (SPS)  & optimization & NB15 & & rate & behaviour \\ \hline 
Raja et al. \cite{Raja2022} & E$\_$SVM-NN & Data Mining & OSN & Python  & Accuracy  & Fake/ \\
(2021) & classifier & tools & accounts & & & accounts \\ \hline
Ranjan et al. \cite{Ranjana2022} & Random  & LAMP stack  & Magneto & MSSQL & Accuracy & Malicious \\
(2022) & forest  & based AWS cloud &  & & & user \\ \hline
Gupta et al. \cite{9865138} & Quantum qubits  & Multilayer feed  & CMU & Python   & Accuracy & Malicious \\
(2022) & with pauli gate  & forward network & CERT & & & user \\ 
\hline
\botrule
\end{tabular*}
\end{sidewaystable}
\section{MAIDS Model}\label{sec3}
The framework for the proposed MAIDS model is manifested collectively by Fig. \ref{fig2} and Fig. \ref{fig3}. Here, Fig. \ref{fig2} demonstrates the operational workflow for the agent eligibility estimation unit (AEE) whereas Fig. \ref{fig3} displays the layout for the XGBoost-based malicious agent prediction unit (XC-MAP). Overall interaction among separate units is reflected in Fig. \ref{op_flow}. Besides, the significant model entities: Agents (${A}$), Data Proprietors (${DP}$), and Cloud Service Provider (\textit{CSP}) are described as follows:
\begin{figure}[!htbp]%
\centering
\includegraphics[width=1.0\textwidth]{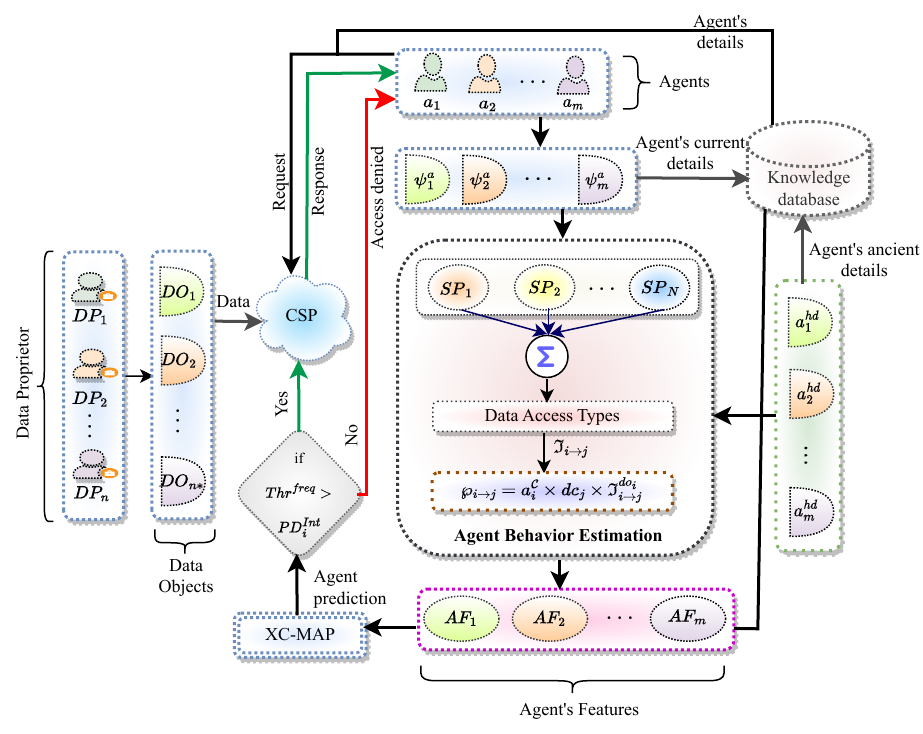}
\caption{Framework of AEE: Agent Eligibility Estimation}\label{fig2}
\end{figure}
\begin{itemize}
    \item \textit{Agents (${A}$)}: An entity that raises data access requests for some utility purpose. It is considered a non-trusted entity in the proposed model. \textit{CSP} scrutinizes the data requests from XC-MAP for any hidden malicious intention before allocating requested data to agents. ${A}$ belongs to one of three vital classes: 'malicious', 'non-malicious', and, 'unknown'.
    \item \textit{Data Proprietors (${DP}$)}: An entity is responsible for data objects \{${DO}_1$, ${DO}_2$, ..., ${DO}_{n^\ast}$\} contribution to the \textit{CSP}. Here, ${DP}$ acts like a non-trusted entity in the proposed model as it might leak data itself.
    \item \textit{Cloud Service Provider (\textit{CSP})}: An entity that receives a data request from distinct agents ${A}$ along with ancient/historical details ($a^{hd}$) and live details (${\psi}^a$), allows data storage, analysis, and sharing. It utilizes the computation ability of AEE (Section \ref{sec4}) and XC-MAP (Section \ref{sec5}) to find out if the data access request is 'malicious' or 'non-malicious'.
\end{itemize}
The proposed study consider $n$ data proprietors: \{${DP}_1$, ${DP}_2$, ..., ${DP}_n$\} $\in {DP}$, sharing $n^\ast$ data objects: \{${DO}_1$, ${DO}_2$, ..., ${DO}_{n^\ast}$\} with $\textit{CSP}$ for storage, computation and sharing among $m$ agents: \{$a_1$, $a_2$, ..., $a_m$\} $\in {A}$ to fulfill their data access request. The proposed model assume that the agents: \{$a_1$, $a_2$, ..., $a_m$\} raises request to $\textit{CSP}$ for data access as reflected in Fig. \ref{fig2}. \textit{Agent's live details}: \{${\psi}_1^a$, ${\psi}_2^a$, ..., ${\psi}_{m}^a$\} along with \textit{agent's historical knowledge}: \{$a^{hd}_1$, $a^{hd}_2$, ..., $a^{hd}_m$) associated to data access requests are supplied into \textit{Agent Eligibility Estimation} (AEE) unit to estimate the agent behavior. It evaluates multiple ($N$) \textit{security risk parameters}: ${SP}$ = \{${SP}_{1}$ $\cup$ ${SP}_2$ $\cup$ ... $\cup$ ${SP}_{N}$\} associated with all $m$ agents request to fetch the \textit{security risk information}: \{$\wp_1$, $\wp_2$, ..., $\wp_m$\}$\in \wp$ (more details in Section \ref{sec4}). Finally, it computes the values of \textit{agent features ($AF$)}: \{$AF_{1}$, $AF_{2}$, $\dots$, $AF_{m}$\} and forwards it to XC-MAP unit for further analysis to accomplish the purpose of proactive malicious agent prediction. Moreover, the agent's live details ($\psi^a$), the agent's historical details ($a^{hd}$), and computed agent features ($AF$) are stored in a knowledge database for model re-training purposes.
\begin{figure}[!htbp]%
\centering
\includegraphics[width=1.0\textwidth]{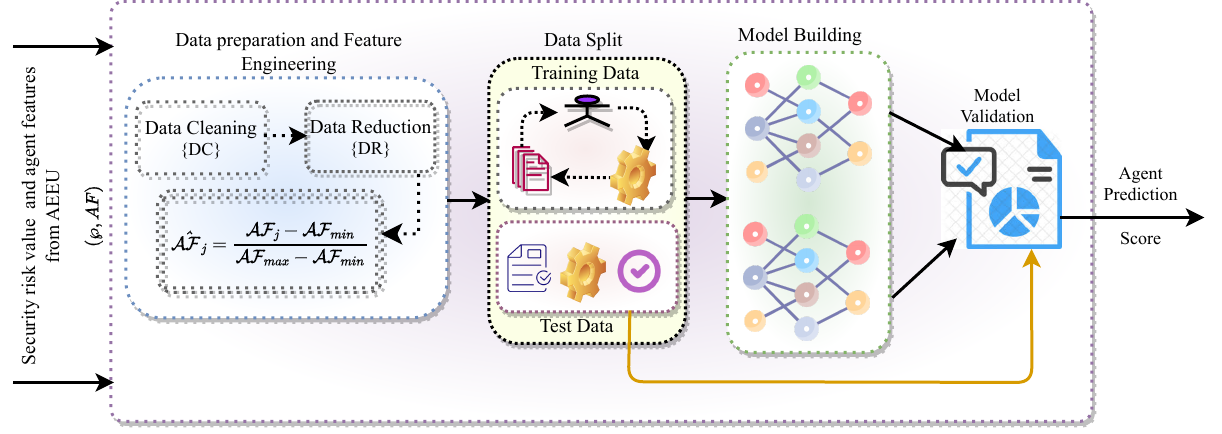}
\caption{Framework of XC-MAP: XGBoost Classification-based Malicious Agent Prediction}\label{fig3}
\end{figure}

As illustrated in Fig. \ref{fig3}, XGBoost classification is utilized to investigate and predict the possibility of malicious activity before the allocation of requested data in a real-time cloud communication environment. This unit takes benefit of the high computational power of XGBoost and machine learning concepts. XC-MAP obtains and examine agents' feature \{$AF_{1}$, $AF_{2}$, $\dots$, $AF_{m}$\} alongwith multiple security risk values \{$\wp_1$, $\wp_2$, ..., $\wp_m$\} $\in$ $\wp$ in respect of various agents received from AEE. This unit is periodically trained and re-trained with updated data samples to predict possible malicious agents with high accuracy. For outcomes determined as non-malicious by XC-MAP; data is distributed among requesting agents keeping high security. The detailed narration of AEE and XC-MAP is explained in the following Section \ref{sec4} and Section \ref{sec5}, respectively.
\section{Agent Eligibility Estimation }\label{sec4}
The agent eligibility for requested data access is computed and analyzed in terms of several security parameters including a total number of \textit{ malicious data distributions or data leakages} (${L}_i^{Int}$) performed by agent $a_i$ over a time-interval \{$t_\alpha$, $t_\beta$\};  \textit{frequency of data breaches} (${PD}^{Int}$); \textit{agent's historical knowledge} ($a^{hd}$); keeping track of each \textit{agent's authorized set of data units} (${PDO}$). The mathematical formulation and estimation of these security parameters are discussed as follows:
\begin{itemize}
    \item \textit{Malicious data distributions} (${L}_i^{Int}$): Suppose the agent $a_{i}$ had demanded data \{$do_{1}$, $do_{2}$, $\dots$, $do_{z}$\} in some previous time-interval \{$t_{\alpha-1}$, $t_{\beta-1}$\} and the total number of unauthorized data distributions (${L}^{Int}$) during this period is estimated using Eq. (\ref{sp4}), wherein $\theta_{i}$ represents leakage status of data such as $\theta_{i} = \{0, 1\}$ which is true (1) or false (0). The attack component details (${ACD}$) for the agent $a_{i}$ is computed using Eq. (\ref{sp5}), where ${DA}^{gross}$ is the total number of data access over period \{$t_{\alpha-1}$, $t_{\beta-1}$\}. Eq. (\ref{sp6}) determines whether an agent $a_i$ should allow access to data or not based on the previous malicious data distributions.
\begin{gather} \label{sp4}
    {L}_i^{Int} = \sum_{i=1}^{z}(do_{i} \times \theta_{i} \times t) \quad  \forall_t \in \{t_{\alpha-1}, t_{\beta-1}\}
\\ \label{sp5}
\int_{t_{\alpha-1}}^{t_{\beta-1}} {ACD}_{i} dt = \int_{t_{\alpha-1}}^{t_{\beta-1}} \frac{{L}_i^{Int}}{{DA}_i^{gross}} dt
\\ \label{sp6}
{SP}_{{ACD}_{i}} = 
 \begin{cases}
  \textit{Access allowed} \,(0),  & If (Thr^{attack} > {ACD}_{i})\\
  \textit{Access denied} \,(1), & Otherwise\\
 \end{cases}
 \end{gather}
\item \textit{Frequency of data breaches} (${PD}^{Int}$): The attempt of unauthorized data leakage probability (${PD}^{Int}$) is computed using Eq. (\ref{sp7}), where $\sum\limits_{k=1}^{H} doz_{k} \not\in {PDO}_{i}$ and $t_{ijk}$ represents number of times the $i^{th}$ agent $a_{i}$ has attempted to access unauthorized data ($doz_{k}$) over $j^{th}$ time-period. The term $H$ and $T$ stands for total number of unauthorized data requested by $a_{i}$ during time duration $T$ where, $T$ $\in$ \{$t_{(\alpha-1)}$, $t_{(\beta-1)}$\}.
Eq. (\ref{sp8}) examines that agent $a_{i}$ should be given data $do_{j}$ only if ${SP}_{{PD}^{Int}}$ = 0, otherwise access is denied, where $Thr^{freq}$ is threshold frequency of illegal data access attempts during period \{$t_{(\alpha-1)}$, $t_{(\beta-1)}$\}.
\begin{gather}
\label{sp7}
    {PD}^{Int}_{i} = |\sum_{k=1}^{H}\sum_{j=1}^{T} doz_{k} \times t_{ijk} \times a_{i}|\\
    \label{sp8}
{SP}_{{PD}^{Int}} = 
 \begin{cases}
\textit{Allowed} \,(0),  & If (Thr^{freq} > {PD}^{Int}_{i})\\
  \textit{Denied} \,(1), & Otherwise\\
 \end{cases}
 \end{gather}
 \item \textit{Agent's historical information} ($a^{hd}$): The agents' history of data access, i.e., $a^{hd}$ is examined using Eq. (\ref{sp1}) to determine whether the agent $a_i$ is `known' or `unknown', wherein ${SP}_{a^{hd}_i}$ signifies security factor associated with $a^{hd}_i$ as follows:
\begin{gather} \label{sp1}
{SP}_{a^{hd}_i} = 
 \begin{cases}
  \textit{Known} \,(0),  & \text{If}  (|a^{hd}_i| > 0)\\
 \textit{Unknown} \,(1), & Otherwise\\
 \end{cases}
 \end{gather}
 \item \textit{Agent's authorized set of data units} (${PDO}$): The agent $a_{i}$ is eligible to access data from the following set of permissible data units (${PDO}_i$) stated in Eq. (\ref{sp2}), wherein $q_{1}$, $q_{2}$, $\dots$, $q_{n^\ast}$ specifies the number of datasets of categories: ${C}_{1}$, ${C}_{2}$, $\dots$, ${C}_{n^\ast}$, respectively, legally allowed to access to $i^{th}$ agent data. Likewise, eligibility for data access is given in Eq. (\ref{sp3}).
\begin{gather}\label{sp2}
{PDO}_i = ({C}_{1} \times \sum_{k=1}^{q_{1}}do_{k}) \cup ({C}_{2} \times \sum_{k=1}^{q_{2}}do_{k}) \cup \dots \cup ({C}_{n^\ast} \times \sum_{k=1}^{q_{n^\ast}}do_{k})
 \\
  \label{sp3}
{SP}_{{PDO}_i} = 
 \begin{cases}
 \textit{Authorized} \,(0),  & \text{If} (a_{i} \times ({C}_{i} \times do_{i}) \subseteq {PDO}_{i})\\
  \textit{Unauthorized} \,(1), & Otherwise\\
 \end{cases}
 \end{gather}
\end{itemize}
Finally, total ${N}$ security factors are aggregated using Eq. (\ref{final1}), and intention of data access is analyzed by applying Eq. (\ref{final2}), where $\Im_{i \rightarrow j}^{do_{i}}$ detects relation or type of access as non-malicious  ($\Im_{i \rightarrow j}^{do_{i}}$ = 0) or malicious  ($\Im_{i \rightarrow j}^{do_{i}}$ = 1) over time-period \{$t_{\alpha}$, $t_{\beta}$\} by applying Eq. (\ref{final3}).
\begin{gather}
\label{final1}
{SP} ={SP}_{a^{hd}_i} +{SP}_{{PDO}_i}+{SP}_{{ACD}_{i}}  +{SP}_{{PD}^{Int}}+ \dots + {SP}_{N}\\
\label{final2}
    \Im_{i \rightarrow j}^{do_{i}}  = 
      \begin{cases}
           \textit{Non-malicious}\,(0),  & If ({SP} < 1)\\
            \textit{Malicious} \,(1), & Otherwise\\
        \end{cases}
        \\ 
        \label{final3}
       \int_{t_{\alpha}}^{t_{\beta}}\wp_{i \rightarrow j} dt = \int_{t_{\alpha}}^{t_{\beta}} (a_{i}^{{C}} \times do_{j} \times \Im_{i \rightarrow j}^{do_{i}}) dt  
        \end{gather}     \\   
\section{Malicious Agent Prediction}\label{sec5}
XC-MAP predicts malicious agents well in advance to ensure high security. The malicious agent prediction model comprises of mainly two main steps. Firstly to prepare the data for processing and next is to carry out the task of objective function optimization as described in the following subsections.
\subsection{Data Preparation}\label{subsec5a}
Cloud Service Provider acquires the distinct agents' details \{$AF_{1}$, $AF_{2}$, $\dots$, $AF_{m}$\} along with multiple security risk values  \{$\wp_1$, $\wp_2$, ..., $\wp_m$\} $\in$ $\wp$ from AEE. Preprocessing is performed to avoid any kind of data mishandling which might lead to an impact on the models' performance. Preprocessing includes steps like data cleaning to take care of missing data (if any), and encoding categorical data by using one hot encoder. \{$AF_{1}$, $AF_{2}$, $\dots$, $AF_{m}$\} details are pre-processed using Eq. (\ref{sp12}).
\begin{gather} \label{sp12}
{\hat{AF}_j} = \frac{{AF}_j - {AF}_{min}}{{AF}_{max} - {AF}_{min}}
\end{gather}
Where ${AF}_{max}$ and ${AF}_{min}$ are the maximum and minimum values of the input data set. In the next step this pre-processed data \{${\hat{AF}}_1$, ${\hat{AF}}_2$, ..., ${\hat{AF}}_m$\} is split into two sets naming training data set and testing data set. The classification model uses a supervised form of learning to train the model. This training data set acts as labeled data and is fed to the model so that the model can learn from it. Thereafter, the testing dataset is used to examine the model performance in the form of parameters like accuracy. Various regularization parameters such as shrinkage parameter ($\lambda$), minimum loss reduction gain ($\gamma$), and min-child-weight are adjusted throughout the model-building process to avoid any kind of overfitting in the XC-MAP model. The $p$ training data samples: \{${\hat{AF}}_1$, ${\hat{AF}}_2$, ..., ${\hat{AF}}_p$\}  $\in$ ${\hat{AF}}$ are drawn out, racked up, normalized, and then finally transformed into an input vector to fed to the underlying model.
\subsection{XC-MAP Optimization}\label{subsec5b}
First of all an initial basic prediction is assumed to fit a training dataset by employing the XGBoost machine learning algorithm. Afterward, on behalf of the predicted value initially and observed value; residuals are evaluated. Thereon, similarity scores are computed for residuals. This leads to the formation of a decision tree. Now, the similarity of the data in a leaf is computed, accompanying the computation of gain in similarity in the subsequent split. A comparison of these gains values is performed to find out a feature and a threshold for the node. Then, for each leaf output value is also evaluated by employing residuals. For the task of agent classification, these values are computed using the log-loss function and probabilistic approaches. Hence, this output of the tree now acts as a new residual for the dataset, which is used further to create more trees. Moreover, the output received from each tree is multiplied by a learning rate parameter added to the initial prediction to compute the final prediction value.\\

Let the model consider having training dataset ${D}$ = $({X}_i, {y_i})^n_i$ with $\textit{n}$ examples \{(${X}_1, {y_1}$), (${X}_2, {y_2}$), ..., (${X}_n, {y_n}$)\} $\in$ ${D}$ and $\textit{m}$ agent features. The ultimate goal is to find an optimization function ${\hat{F}}({X})$ for the objective function, to map the input ${X}$ into its output values $y$ as stated in Eq. (\ref{sp13}) by minimizing the expected values of the Loss function, $\zeta(y,{F}({X}))$.
\begin{gather}
\label{sp13}
\hat{y}_i = {F}({{X}_i}, {y_i})
\end{gather}
\begin{itemize}
    \item \textit{Objective optimization}: Gradient Boosting trains the model in an additive manner to build an additive approximation of the objective function. Consider that the $\hat{y}^{(t)}_i$ is the prediction outcome for $i^{th}$ instance at $t^{th}$ iteration. Therefore $f_t$ is added to minimize the objective function as given in Eq. (\ref{sp14}). A log-loss differential loss function ($l$) measures the difference between predicted $\hat{y}_i$ and actual $y_i$ as given in Eq. (\ref{sp14a}) where a sigmoidal function is used to compute the value of $p$ as shown in Eq. (\ref{sp14b}). 
    Here, Eq. (\ref{sp15}) defines \textbf{$\Omega$}, which is the penalization factor to limit the complexity of the proposed model. 
    \begin{gather}
    \label{sp14}
    \zeta^{(t)} = \sum_{i=1}^{n} l(y_i, \hat{y}^{(t-1)}_i + f_t({X}_i))+\Omega(f_t) \\
        \label{sp14a}
        l = [-y\log(p) + (1-y)\log(1-p)] \\
        \label{sp14b}
        p = \frac{1}{1 + e^{-\hat{y}_i}} \\
        \label{sp15}
        \Omega(f) = \gamma{T} + \frac{1}{2}\lambda{||w||^2}
    \end{gather}

Several regularisation parameters are used to avoid model over-fitting and faster model training with less storage space requirements. Here, \textbf{$\gamma$} controls the minimum loss reduction gain to split an internal node. A higher $\gamma$ value leads to the simple tree. It boosts training and minimizes storage requirements. \textbf{$\lambda$} act as shrinkage factor; the step size (learning rate) to reduce the influence of individual trees and create a way for future trees to improve the model performance. To have a deep insight into model performance the underlying model is also trained and tested for varying weight values; shown in section \ref{subsec7b}. It is for another regularisation parameter to avoid over-fitting in the model. \textbf{${T}$} represents the total number of leaf nodes for a tree and $w$ is the weight/output score of leave. \\
 \item \textit{Loss estimation}: The proposed XGBoost approach uses the \textit{Taylor expansion series} as represented in Eq. (\ref{sp16}), to compute the values of the loss function already defined in Eq. (\ref{sp14}), for a base learner $f_t({X}_i)$. Taylor series can be expanded up to numerous derivative orders, but in the proposed model, the Taylor expansion up to second-order expanded terms is utilized to obtain an estimated approximated loss value. The major reason for this is that an up-to-two-level approximation value is good enough to serve the purpose without causing much computation and complexity. Here, $g_if_t$ and $h_if^2_i$ are the first-order and second-order derivatives, represented computed in Eq. (\ref{sp17}) and Eq. (\ref{sp18}), respectively. The remaining terms are dropped off for the ease of computation of the proposed approach without causing much variation from actual results. 
    \begin{gather}
    \label{sp16}
    \zeta^{(t)} \approx \sum_{i=1}^{n} l[(y_i, \hat{y}^{(t-1)}_i) + g_if_t({X}_i) + \frac{1}{2}h_if^2_t({X}_i)]+\Omega(f_t) \\
        \label{sp17}
        g_i = \partial_{\hat{y}^{(t-1)}}l(y_i, \hat{y}^{(t-1)}_i ) \\ 
        \label{sp18}
        h_i = \partial^2_{\hat{y}^{(t-1)}}l(y_i, \hat{y}^{(t-1)}_i ) 
    \end{gather} \\
    \item \textit{Similarity score evaluation}: Simplified objective function shown in Eq. (\ref{sp19}) is obtained by eliminating constant term. Optimal weight $w^*_j$ for a leaf and optimal value for a tree structure $q(x)$ are obtained from Eq. (\ref{sp20}) and Eq. (\ref{sp21}) respectively. $\Bar{\zeta}^{(t)}(q)$ specifies the quality of tree structure $q$.
    \begin{gather}
    \label{sp19}
    \Bar{\zeta}^{(t)} = \sum_{i=1}^{n} [g_if_t({X}_i) + \frac{1}{2}h_if^2_t({X}_i)]+\Omega(f_t) \\
    \label{sp20}
    w^*_j = \frac{\sum_{i \in I_j} g_i }{\sum_{i \in I_j} h_i + \lambda} \\
    \label{sp21}
    \Bar{\zeta}^{(t)}(q) = -\frac{1}{2}\sum_{j=1}^{T}\frac{(\sum_{i \in I_j} g_i )^2}{\sum_{i \in I_j} h_i + \lambda} + \gamma{T}
    \end{gather}

XGBoost is an exactly greedy algorithm. It starts from a single leaf node and keeps on growing by adding branches either on the left side or right side. Residuals are computed based on predicted value and observed as $\textit{Residual= Observed values- Actual values}$. A split is performed into $I_L$ and $I_R$; the instance of left tree and right tree, here $I = I_L \cup I_R$. \\
    \item \textit{Gain computation}: To obtain the overall gain from this tree structure $q$ after splitting into branches ($I_L, I_R)$ can be calculated as given in Eq. (\ref{sp22}).
    \begin{gather}
    \label{sp22}
    \zeta_{split} = \frac{1}{2}[\frac{(\sum_{i \in I_L} g_i )^2}{\sum_{i \in I_L} h_i + \lambda} + \frac{(\sum_{i \in I_L} g_i )^2}{\sum_{i \in I_R} h_i + \lambda} - \frac{(\sum_{i \in I} g_i )^2}{\sum_{i \in I} h_i + \lambda}] - \gamma
    \end{gather}
\end{itemize}

Being an exact greedy algorithm XGBoost builds a tree by exploring all possible splits, that too for all the features. Model keep trying to grow for all the possible tree structures but it's a very tedious and sort of impossible task to explore all the possible tree structure $q$, therefore the split which imparts maximum loss reduction is chosen. It is considered that the gain for the best split must be positive and must be greater than the min-split-gain parameter otherwise, stop growing the branch further. This yields the split with the maximum score. Based on this score XC-MAP model predicts the possibility of any agent being 'malicious' or 'non-malicious.'
\section{Operational Design and Complexity Computation}\label{sec6}
Algorithm \ref{algo1} conveys the entire working of the proposed MAIDS in a very crisp form. This algorithm successfully transmits the fact that the MAIDS model is operating proactively to ensure highly secure data communication over the cloud platform. Firstly, the agents, their attributes, and data requests are initialized. In the next step XC-MAP model training and re-training are performed periodically. Steps 3-16 iterate for $\textit{t}$ intervals to determine whether to grant access or not to the requesting agent for the requested set of data.
\begin{algorithm}
\caption{MAIDS: Operational Layout}\label{algo1}
\begin{algorithmic}[1]
\STATE Initialize: agents list ($List_{{A}}$) with associated features and requested data\;
\STATE XC-MAP training and re-training periodically, with live and historical malicious agent data instance \;
\FOR{each time-interval \{$t_{\alpha}$, $t_{\beta}$\}}
    \FOR{each agent ($a_i: i \in$ \{1, 2, ..., m\})}
        \STATE Receive requests from agent and analyse \\
          $ {SP}$=\{${SP}_1 \cup {SP}_2 \cup ...\cup {SP}_{N}$\}\;
        \STATE Examine the possible purpose related to data access request by evaluating Eq. (\ref{final3})\; \\
        \STATE $\wp^{Int} \Leftarrow$ {XC-MAP($\wp$,\{${AF}_1$, ${AF}_2$, ..., ${AF}_m$\})}\; as computed in Algorithm \ref{algo2}
        \IF{$\wp^{Int} $ $>$ $0$}
            \STATE {Agent $a_i$ is `Malicious '}
            \STATE {Access to data not granted \;}\;
        \ELSE
            \STATE {Access to data is granted\;}\;
            \STATE {Distribute requested data ${DO}_i$ to agent}
            \ENDIF
    \ENDFOR
\ENDFOR
\end{algorithmic}
\end{algorithm}

Algorithm \ref{algo2} showcases the working pathway of XC-MAP: XGBoost machine learning classifier-based malicious agent prediction unit in a summarised form. First of all the current node instance is considered thereafter various agent features ($AF$) and security risk information ($\wp$) received from the agent eligibility estimation unit are fed as input into XC-MAP to carry out agent analysis further. In steps (3-4) basic are performed and in steps (5-12) iteration is repeated for all agents. Every iteration computes the score value, residual, and gain for the left and right child for each instance of the tree. Step 13 computes the split with maximum value, the final output from the model.
\begin{algorithm}
\caption{XC-MAP: Working Pathway}\label{algo2}
\begin{algorithmic}[1]
\STATE Input: \textbf{C} instance of current node\;
\STATE Input: agent features and security parameters ($\wp$,\{${AF}_1$, ${AF}_2$, ..., ${AF}_m$\})\; \\
\STATE Initialize: gain = 0
\STATE $G = \sum_{i\in C}g_i $ and $ H = \sum_{i\in C}h_i$\; \\
\FOR{$k = \{1, 2, \dots, m\}$ }
    \STATE $G_i = 0 $ and $ H_i =  0$ using Eq. (\ref{sp13}) to Eq. (\ref{sp22}) \; \\
    \FOR{j in sorted(C, by $X_{jk}$)}
        \STATE $G_L = G_L + g_i$ , $H_L = H_L + h_i$ \; \\
         \STATE $G_R = G - G_L$ , $H_R = H - H_L$ \; \\ 
        \STATE Score = max (score, $\frac{G^2_L}{H_L + \lambda}$ + $\frac{G^2_R}{H_R + \lambda}$ - $\frac{G^2}{H + \lambda}$) \;
    \ENDFOR
\ENDFOR
\STATE Output: split with maximum score value
\end{algorithmic}
\end{algorithm}
\subsection{Complexity Computation}\label{subsec6a}
Algorithm \ref{algo1} and Algorithm \ref{algo2} together illustrate the proposed model's overall working. In Algorithm \ref{algo1} various operations like agent list initialization and data request initialization are defined in step 1 which causes ${O}(1)$ complexity. Step 2 carries the XC-MAP training and re-training task bearing complexity as computed in Algorithm \ref{algo2}. Steps (3-16) iterate over $t$ intervals, wherein steps (4-15) repeat for $m$ agents. In step 5, $N$ security parameters (${SP}$) for each agent are analyzed while in steps 6 \& 7 intentions of agents for being malicious or non-malicious are examined, imparting complexity ${O}(N)$. Depending on the data request type and predicted agent information approval or rejection of data access is carried out in steps (8-14) owing ${O}(1)$ complexity.

While Algorithm \ref{algo2} accomplishes the XC-MAP training and re-training task periodically by calculating the score values, residuals values and, gain values for each node, left and right branches separately. Here, complexity computation depends on the number of training samples ($x$) and the dimension of the feature ($y$) whose complexity is ${O}(x \times y)$. Hence, the total complexity yields to be ${O}(t \times m \times x \times y \times N ) \Rightarrow$  ${O}(tmxyN)$.
\subsection{Illustration}\label{subsec6b}
Let's assume that the proposed model comprises five agents \{$a_1$, $a_2$, $a_{3}$, $a_{4}$, $a_{5}$\} which have raised requests for data objects \{${DO}_3$, ${DO}_7$, ${DO}_{21}$, ${DO}_{5}$, ${DO}_9$, ${DO}_{24}$\}. It is assumed that the agents, $a_2$ and $a_5$ are malicious and are looking for confidential data (${DO}_{21}$, ${DO}_{24}$) with the intent to breach this data. The requested data is allocated among requesting agents using present data distribution approaches like \cite{5487521}, \cite{gupta2020mlpam}, without considering the agent's intentions about data utility. This might lead to granting of unauthorized data access to the dataset ${DO}_{21}$, ${DO}_{24}$ and later on detect a data breach, and hence now the hunt starts to identify the malicious agent. Instead, the proposed MAIDS model acting proactively, scrutinize the agents along with their features supplied during request raised for data access. The model categorizes the intentions for data utility of every agent and hence comes to decide that the agents: \{$a_1$,  $a_{3}$, $a_{4}$\} are non-malicious whereas the agents: \{$a_2$, $a_{5}$\} are malicious. Consequently, it is observed that MAIDS is contributing a lot to ensure data security and communication by utilizing proactive prediction of possible data leakages even before they happen and also not granting unauthorized data access to sensitive data.
\section{Performance Evaluation and Discussion}\label{sec7}
\subsection{Experimental Set-up}\label{subsec7a}
The experimental work is carried out on a server machine encompassing two Intel\textsuperscript{\textregistered} Xeon\textsuperscript{\textregistered} Silver 4114 CPU with a 40-core processor and having 2.20 GHz clock speed. The simulation machine runs on Ubuntu 16.04, a 64-bit LTS operating system comprising 128 GB of main memory RAM. Enactment of the proposed work is carried out using Python 3.9.
\subsection{Implementation}\label{subsec6c}
Fig. \ref{implementation} depicts the design of the proposed model. Specifically, the MAIDS model is shaped with the collaboration of the different modules as discussed below:
\begin{figure}[!htbp]%
\centering
\includegraphics[width=0.95\textwidth]{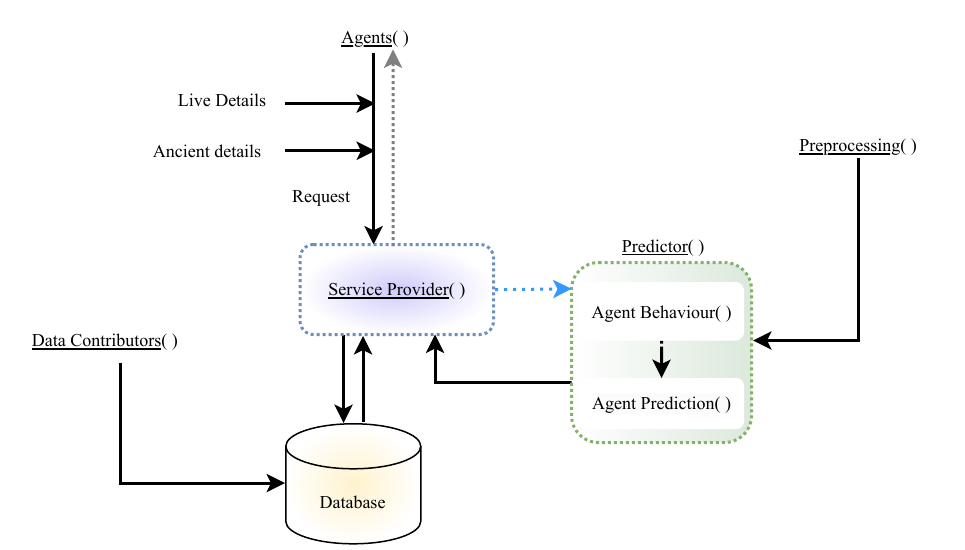}
\caption{Implementation flow for proposed MAIDS Model}\label{implementation}
\end{figure}
\begin{itemize}
    \item \textit{Agents()}: Agents raise a request to grant access to the data from the database. Agents supply live details along with ancient/historical information, to the service provider().
    \item \textit{Data Contributors()}: The contributors are the data proprietors who own the data. The contribution of data to the database is a continuous ongoing process.
    \item \textit{Service Provider()}: This module handles multiple tasks as it receives data requests from agents, can access the database directly, and supports the module to carry out the prediction task.
    \item \textit{Preprocessing()}: The relevant numerical details of agents along with historical and live details are extracted and normalized to prepare input values for training and retraining of the agent prediction model.
    \item \textit{Predictor()}: This module is the soul of model. It comprises two sub-modules to find out agent behavior and predict its intention.
    \item \textit{Agent Behaviour()}: The varying number of agents as mentioned in Section \ref{sec4} are examined to estimate their behavior based on live details and ancient details.
    \item \textit{Agent Prediction()}: The module employs XGBoost machine learning classification as elaborated in Section \ref{sec5} to depict the intention of the agent for granting or denying access to the requested data.
\end{itemize}
\subsection{Dataset}
The performance of the model under consideration is examined through a dataset comprising numerous agents' live details along with ancient details. Major live details parameters are the type of profession, number of requests from the agent, type of requests from the agent, and data limit for which data was accessed. The major ancient details parameters are ancient data of agents, leaked or never leaked data, how many times leaked the data, how frequently asked for data, and data retention. An extended version of CMU CERT synthetic insider threat dataset r4.2 \cite{dataset} is employed for research. Here a scenario with 10k agents is considered. These agents all together are classified into three strictly different brackets which are non-malicious, malicious, and unknown. Moreover, the model assumes all the entities as non trusted to carry out execution tasks.
\subsection{Prediction Accuracy Analysis and Comparison}\label{subsec7b}
\begin{figure}[!htbp]%
\centering
 \includegraphics[height=7cm, width=0.9\textwidth]{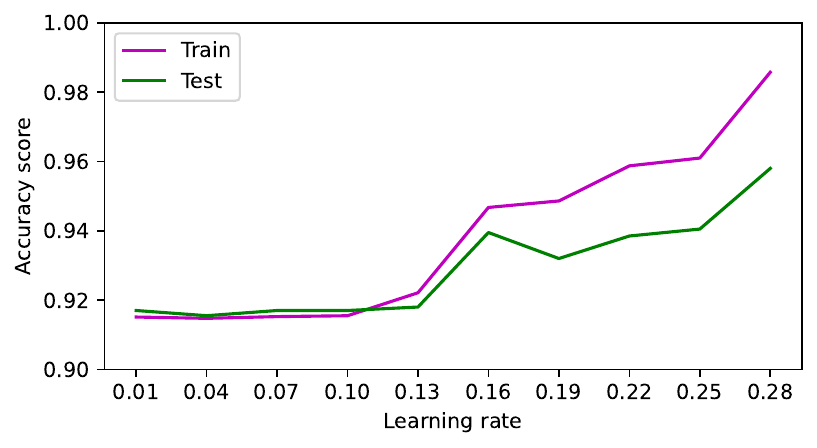}
\caption{Accuracy score v/s Learning rate of Model}\label{figacclearn}
\end{figure}

MAIDS model is comprised of two units AEE: to analyze the agent behavior and XC-MAP: to predict the agent intention, respectively. Fig. \ref{figacclearn} depicts the XC-MAP module accuracy score for the training and testing phase separately over different learning rates. Here observation derived is that the model is performing well enough without causing any over-fitting.
\begin{figure}[ht!]%
    \begin{center}
        \subfigure[$\omega$ = 2.0]{
        \includegraphics[height=5cm, width=0.9\textwidth]{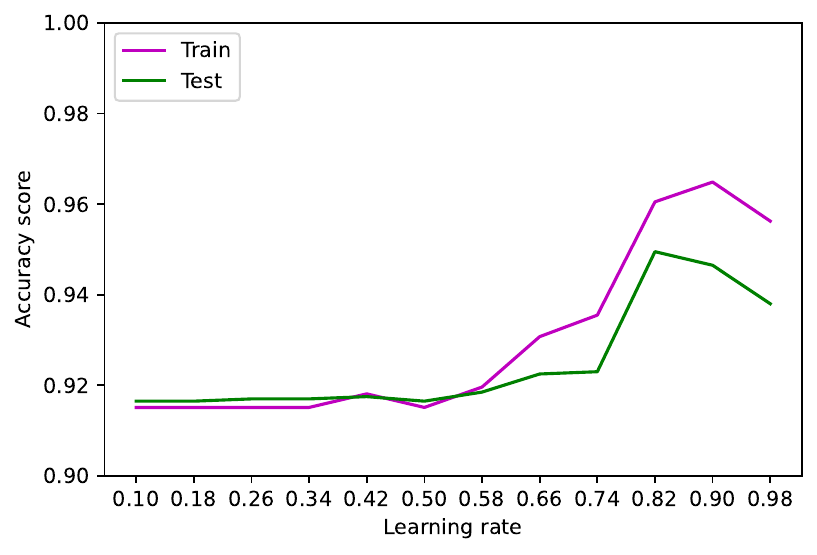}
        }%
        \\
        \subfigure[$\omega$ =  3.0]{
        \includegraphics[height=5cm, width=0.9\textwidth]{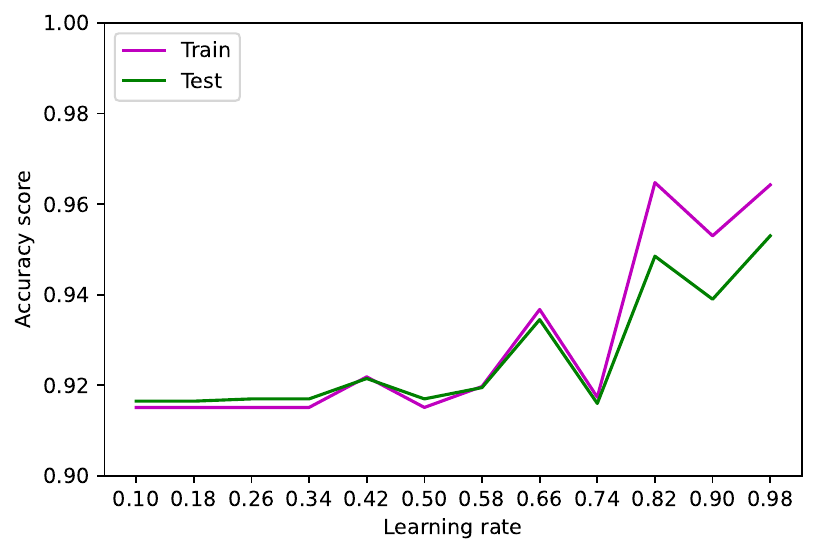}
        }%
        \\
        \subfigure[$\omega$ =  4.0]{
        \includegraphics[height=5cm, width=0.9\textwidth]{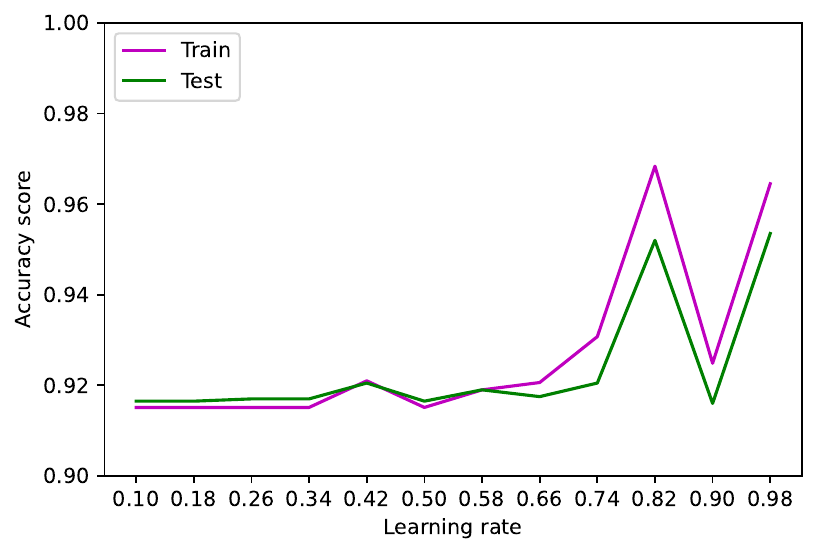}
        }%
    \end{center}       
\caption{Accuracy score v/s Learning rate of model for different $\omega$ }\label{figacclearnweight}
\end{figure}

Fig. \ref{figacclearnweight} presents a more detailed picture of the model performance parameter accuracy score and learning rate in the range zero to one that too for different weights (2.0, 3.0, 4.0) during the model training phase and testing phase. These figures demonstrate the efficiency and smooth functioning of the model without any overfitting of agent features in the model.
\subsection{Comparison}
A comparison of the MAIDS model is carried out with various already proposed state-of-the-art approaches such as \textit{Attribute/Behavior-Based Access Control} (ABBAC) scheme \cite{afshar2021incorporating}, \textit{Guilty Agent Model} (GAM) \cite{5487521},  \textit{Dynamic Threshold based Information Leaker Identification scheme} (DT-ILIS) \cite{gupta2019dynamic}, \textit{Machine Learning and Probabilistic Analysis Based Model} (MLPAM) \cite{gupta2020mlpam}, and \textit{Quantum Machine Learning driven Malicious User Prediction Model} (QM-MUP) \cite{9865138}.

ABBAC established an access control approach by utilizing various user behavior details to find malicious entities. GAM employed the concept of data allocation with the principle of minimum overlapping to find the malicious entity based on the data allocation pattern. DT-ILIS proposed an approach to identify the data leaker depending on the preset threshold value. MLPAM presented a sophisticated guilty agent prediction strategy using data distribution and probability metrics. QM-MUP incorporated an all-in-first-time approach to identify guilty ones proactively by deploying the computability of quantum mechanics in the form of qubits and quantum Pauli gates.

MAIDS devised an approach for malicious agent identification in a proactive manner by utilizing the different agent behavior analysis along with the computation efficiency of XGBoost, a machine learning classification tool. It computes the malicious entity identification task with much higher efficiency and high value of various performance parameters.
 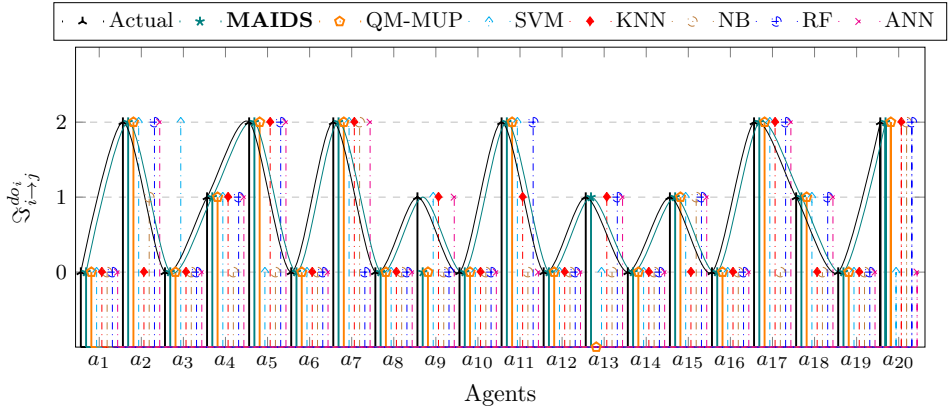
\begin{figure*}[!htbp]%
\begin{center}
\begin{tikzpicture}[node distance = 1cm,auto,scale=.90, transform shape]
\pgfplotsset{every axis y label/.append style={rotate=180,yshift=10.5cm}}
\begin{axis}[
      axis on top=false,
      width=14cm,
        height=6.0cm,
      xmin=15, xmax=338,
      ymin=0, ymax=4,
      ytick={1,2,3},
      yticklabels={\scriptsize 0,\scriptsize 1,\scriptsize 2},
      xtick={24,40,56,72,88,104,120,136,152,168,184,200,216,232,248,264,280,296,312,328},
      xticklabels={\scriptsize $a_{1}$,\scriptsize $a_{2}$,\scriptsize $a_{3}$, \scriptsize $a_{4}$,\scriptsize $a_{5}$, \scriptsize $a_{6}$, \scriptsize $a_{7}$, \scriptsize $a_{8}$, \scriptsize $a_{9}$, \scriptsize $a_{10}$, \scriptsize $a_{11}$, \scriptsize $a_{12}$, \scriptsize $a_{13}$,\scriptsize $a_{14}$,\scriptsize $a_{15}$, \scriptsize $a_{16}$, \scriptsize $a_{17}$, \scriptsize $a_{18}$,\scriptsize $a_{19}$,\scriptsize $a_{20}$},
        ycomb,
        ylabel near ticks, yticklabel pos=left,
      ylabel={$\Im_{i \rightarrow j}^{do_{i}}$},
      xlabel={Agents},
      legend style={at={(0.5,1.15)},
      anchor=north,legend columns=8},
      ymajorgrids=true,
      grid style=dashed,
          ]
\addplot+[mark=Mercedes star,mark options={fill=black},fill=black,draw=black,thick] 
coordinates
{(17,1) (33,3) (49,1) (65,2) (81,3) (97,1) (113,3) (129,1) (145,2) (161,1) (177,3) (193,1) (209,2) (225,1) (241,2) (257,1) (273,3) (289,2) (305,1) (321,3)}
\closedcycle;%
\addlegendentry{\small Actual}
\addplot+[mark=star,mark options={fill=teal},fill=teal,draw=teal,thick] 
coordinates
{(19,1) (35,3) (51,1) (67,2) (83,3) (99,1) (115,3) (131,1) (147,1) (163,1) (179,3) (195,1) (211,2) (227,1) (243,2) (259,1) (275,3) (291,2) (307,1) (323,3)}
\closedcycle;%
\addlegendentry{\small \textbf{MAIDS}}
\addplot+[mark=pentagon,mark options={fill=orange},fill=orange,draw=orange,thick] 
coordinates
 {(21,1) (37,3) (53,1) (69,2) (85,3) (101,1) (117,3) (133,1) (149,1) (165,1) (181,3) (197,1) (213,0) (229,1) (245,2) (261,1) (277,3) (293,2) (309,1) (325,3)}
\closedcycle;
\addlegendentry{\small {QM${\text -}$MUP}}
\addplot+[mark=diamond,mark options={fill=cyan},fill=cyan,draw=cyan,thick, thin,dash dot] 
coordinates
 {(23,1) (39,3) (55,3) (71,2) (87,1) (103,1) (119,3) (135,1) (151,2) (167,1) (183,3) (199,1) (215,1) (231,1) (247,2) (263,1) (279,3) (295,2) (311,1) (327,1)}
\closedcycle;
\addlegendentry{\small SVM}
\addplot+[mark options={fill=red},fill=red,draw=red,thick, thin,dash dot] 
coordinates
 {(25,1) (41,1) (57,1) (73,2) (89,3) (105,1) (121,3) (137,1) (153,2) (169,1) (185,2) (201,1) (217,2) (233,1) (249,1) (265,1) (281,3) (297,1) (313,1) (329,3)}
\closedcycle;
\addlegendentry{\small KNN}
\addplot+[mark=o,mark options={fill=brown},fill=brown,draw=brown,thick, thin,dash dot] 
coordinates
 {(27,1) (43,2) (59,1) (75,1) (91,1) (107,1) (123,3) (139,1) (155,1) (171,1) (187,1) (203,1) (219,1) (235,1) (251,2) (267,1) (283,1) (299,1) (315,1) (331,3)}
\closedcycle;
\addlegendentry{\small NB}
\addplot+[mark=oplus,mark options={fill=blue},fill=blue,draw=blue, thick, thin,dash dot] 
coordinates
 {(29,1) (45,3) (61,1) (77,2) (93,3) (109,1) (125,1) (141,1) (157,1) (173,1) (189,3) (205,1) (221,2) (237,1) (253,2) (269,1) (285,3) (301,2) (317,1) (333,3)}
\closedcycle;
\addlegendentry{\small RF}
\addplot+[mark=x,mark options={fill=magenta},fill=magenta,draw=magenta,  thick, thin,dash dot] 
coordinates
 {(31,1) (47,3) (63,1) (79,2) (95,3) (111,1) (127,3) (143,1) (159,2) (175,1) (191,1) (207,1) (223,2) (239,1) (255,2) (271,1) (287,3) (303,2) (319,1) (335,1)}
\closedcycle;
\addlegendentry{\small ANN}
   \addplot[line width=0mm,draw=black,smooth] 
       coordinates {(17,1) (33,3) (49,1) (65,2) (81,3) (97,1) (113,3) (129,1) (145,2) (161,1) (177,3) (193,1) (209,2) (225,1) (241,2) (257,1) (273,3) (289,2) (305,1) (321,3)};
   \addplot[line width=0mm,draw=teal,smooth] 
       coordinates {(19,1) (35,3) (51,1) (67,2) (83,3) (99,1) (115,3) (131,1) (147,2) (163,1) (179,3) (195,1) (211,2) (227,1) (243,2) (259,1) (275,3) (291,2) (307,1) (323,3)};
\end{axis}
\end{tikzpicture}
\caption{Malicious Agent: Actual v/s Predicted Behavior}\label{fig20agent}
\end{center}
\end{figure*}

An elaborated comparison of the actual agent behavior with predicted agent behavior ($\Im_{i \rightarrow j}^{d_{i}}$) by MAIDS and other state-of-the-art methods like QM-MUP \cite{9865138}, K-Nearest Neighbor (K-NN) \cite{afshar2021incorporating}, Random Forest (RF) \cite{afshar2021incorporating}, Support Vector Machine (SVM), Naive Bayes (NB), and Artificial Neural Network (ANN) are depicted in Fig. \ref{fig20agent} by utilizing a very crisp sample set of only 20 agents. It is visible from the figure that the proposed MAIDS model is capable of predicting the agent behavior almost nearly to the actual behavior of the agent. It yields efficiency because of the computational efficiency of XGBoost over existing traditional approaches for the same purpose.
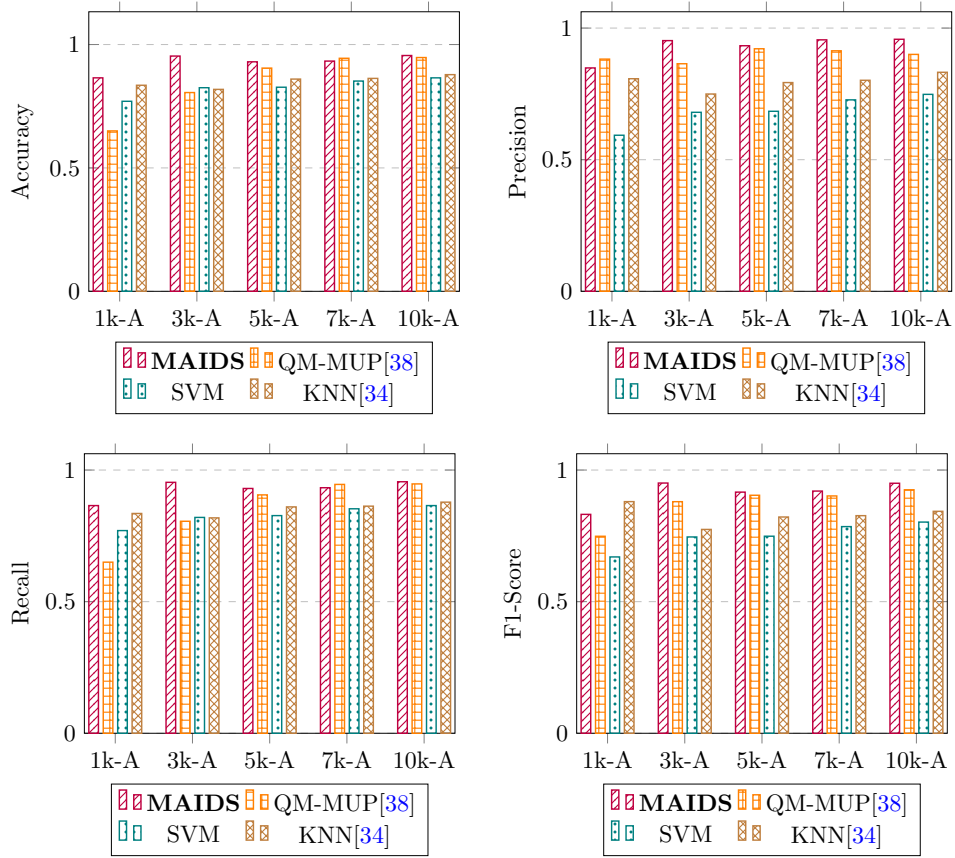
\begin{figure*}[!htbp]
\subfigure{
\begin{tikzpicture}[node distance = 1cm,auto,scale=.90, transform shape]
   \pgfplotsset{every axis y label/.append style={yshift=0cm}}
\begin{axis}[
      axis on top=false,
        height=5.7cm,width=7cm, 
    ybar,      
     bar width=4pt,
    ymin=0,
    ymax=0.96,
  enlarge y limits={upper,value=0.18},
    legend style={at={(0.5,-0.18)},
      anchor=north,legend columns=2},
    ylabel={Accuracy},
  symbolic x coords={1k-A, 3k-A, 5k-A, 7k-A, 10k-A},
    xtick=data,
    nodes near coords align={vertical},
    ymajorgrids=true,
      grid style=dashed,
    ]
\addplot [draw = purple,
    semithick,
    pattern = north east lines,
    pattern color = purple
] coordinates {(1k-A,.8650) (3k-A,.9533) (5k-A,.9300) (7k-A,.9329) (10k-A,.9555) };
\addplot [draw = orange,
    semithick,
    pattern = grid,
    pattern color = orange
] coordinates {(1k-A,.6500) (3k-A,.8054) (5k-A,.9045)  (7k-A,.9443)(10k-A,.9473)};
\addplot [draw = teal,
    semithick,
    pattern = dots,
    pattern color = teal
] coordinates {(1k-A,.7700) (3k-A,.8250) (5k-A,.8270) (7k-A,.8520) (10k-A,.8650)};
\addplot [draw = brown,
     semithick,
     pattern = crosshatch,
     pattern color = brown
 ] coordinates {(1k-A,.8350) (3k-A,.8183) (5k-A,.8600) (7k-A,.8628) (10k-A,.8780)};
\legend{\textbf{MAIDS}, QM-MUP\cite{9865138}, SVM, KNN\cite{afshar2021incorporating}}
\end{axis}
\end{tikzpicture}}%
\hspace{0.2cm}
\subfigure{
\begin{tikzpicture}[node distance = 1cm,auto,scale=.90, transform shape]
   \pgfplotsset{every axis y label/.append style={yshift=0cm}}
\begin{axis}[
      axis on top=false,
        height=5.7cm,width=7cm, 
    ybar,      
     bar width=4pt,
    ymin=0,
    ymax=0.9,
  enlarge y limits={upper,value=0.18},
    legend style={at={(0.5,-0.18)},
      anchor=north,legend columns=2},
    ylabel={Precision},
  symbolic x coords={1k-A, 3k-A, 5k-A, 7k-A, 10k-A},
    xtick=data,
    nodes near coords align={vertical},
    ymajorgrids=true,
    grid style=dashed,
    ]
\addplot [draw = purple,
    semithick,
    pattern = north east lines,
    pattern color = purple
] coordinates {(1k-A,.8486) (3k-A,.9524) (5k-A,.9329) (7k-A,.9553) (10k-A,.9575) };
\addplot [draw = orange,
    semithick,
    pattern = grid,
    pattern color = orange
] coordinates {(1k-A,.8821) (3k-A,.8645) (5k-A,.9214)  (7k-A,.9137)(10k-A,.9000)};
\addplot [draw = teal,
    semithick,
    pattern = dots,
    pattern color = teal
] coordinates {(1k-A,.5929) (3k-A,.6800) (5k-A,.6839) (7k-A,.7273) (10k-A,.7482)};
\addplot [draw = brown,
     semithick,
     pattern = crosshatch,
     pattern color = brown
 ] coordinates {(1k-A,.8076) (3k-A,.7495) (5k-A,.7930) (7k-A,.8017) (10k-A,.8321)};
\legend{\textbf{MAIDS}, QM-MUP\cite{9865138}, SVM, KNN\cite{afshar2021incorporating}}
\end{axis}
\end{tikzpicture}}%
\\
\subfigure{
    \begin{tikzpicture}[node distance = 1cm,auto,scale=.90, transform shape]
   \pgfplotsset{every axis y label/.append style={yshift=0cm}}
\begin{axis}[
    axis on top=false,
    height=5.7cm,width=7cm, 
    ybar,      
     bar width=4pt,
    ymin=0,
    ymax=0.9,
  enlarge y limits={upper,value=0.18},
    legend style={at={(0.5,-0.18)},
    anchor=north,legend columns=2},
    ylabel near ticks, yticklabel pos=left, 
    ylabel={Recall},
    symbolic x coords={1k-A, 3k-A, 5k-A, 7k-A, 10k-A},
    xtick=data,
    nodes near coords align={vertical},
    ymajorgrids=true,
     grid style=dashed,
    ]
\addplot [draw = purple,
    semithick,
    pattern = north east lines,
    pattern color = purple
] coordinates {(1k-A,.8650) (3k-A,.9533) (5k-A,.9300) (7k-A,.9328) (10k-A,.9555) };
\addplot [draw = orange,
    semithick,
    pattern = grid,
    pattern color = orange
] coordinates {(1k-A,.6500) (3k-A,.8054) (5k-A,.9056)  (7k-A,.9454)(10k-A,.9473)};
\addplot [draw = teal,
    semithick,
    pattern = dots,
    pattern color = teal
] coordinates {(1k-A,.7700) (3k-A,.8200) (5k-A,.8269) (7k-A,.8528) (10k-A,.8650)};
\addplot [draw = brown,
     semithick,
     pattern = crosshatch,
     pattern color = brown
 ] coordinates {(1k-A,.8350) (3k-A,.8183) (5k-A,.8600) (7k-A,.8628) (10k-A,.8780)};
\legend{\textbf{MAIDS}, QM-MUP\cite{9865138}, SVM, KNN\cite{afshar2021incorporating}}
\end{axis}
\end{tikzpicture}}%
\hspace{0.2cm}
\subfigure{
    \begin{tikzpicture}[node distance = 1cm,auto,scale=.90, transform shape]
   \pgfplotsset{every axis y label/.append style={yshift=0cm}}
\begin{axis}[
    axis on top=false,
    height=5.7cm,width=7cm, 
    ybar,      
     bar width=4pt,
    ymin=0,
    ymax=0.9,
  enlarge y limits={upper,value=0.18},
    legend style={at={(0.5,-0.18)},
      anchor=north,legend columns=2},
    ylabel={F1-Score},
  symbolic x coords={1k-A, 3k-A, 5k-A, 7k-A, 10k-A},
    xtick=data,
    nodes near coords align={vertical},
    ymajorgrids=true,
     grid style=dashed,
    ]
\addplot [draw = purple,
    semithick,
    pattern = north east lines,
    pattern color = purple
] coordinates {(1k-A,.8316) (3k-A,.9506) (5k-A,.9161) (7k-A,.9201) (10k-A,.9498) };
\addplot [draw = orange,
    semithick,
    pattern = grid,
    pattern color = orange
] coordinates {(1k-A,.7484) (3k-A,.8798) (5k-A,.9042)  (7k-A,.9016)(10k-A,.9250)};
\addplot [draw = teal,
    semithick,
    pattern = dots,
    pattern color = teal
] coordinates {(1k-A,.6699) (3k-A,.7458) (5k-A,.7486) (7k-A,.7851) (10k-A,.8023)};
\addplot [draw = brown,
     semithick,
     pattern = crosshatch,
     pattern color = brown
 ] coordinates {(1k-A,.8800) (3k-A,.7744) (5k-A,.8216) (7k-A,.8267) (10k-A,.8432)};
\legend{\textbf{MAIDS}, QM-MUP\cite{9865138}, SVM, KNN\cite{afshar2021incorporating}}
\end{axis}
\end{tikzpicture}}%
\caption[Optional caption for list of figures]{Comparison of MAIDS's performance parameters for varying Agents with state-of-the-art methods}
\label{figbarchart}
\end{figure*}

In Fig. \ref{figbarchart} and Table \ref{tabdiffagent}, a brief comparison of MAIDS with state-of-the-art works for different numbers of agents for different performance parameters Accuracy, Recall, Precision, and F1-score is presented. It shows a comparative analysis of all performance parameters considering five different scenarios with varying numbers of agents from 1k to 10k in each instance. It highlights that the MAIDS is performing the assigned malicious agent identification task with the highest efficiency and minimum computational complexity in comparison to other methods. Moreover, MAIDS is performing identification tasks proactively rather than after the occurrence of a data leakage event. Hence, MAIDS stands ahead of all the state-of-the-art methods in all manner to ensure data security through data allocation. Therefore, it stands fit for an efficient approach towards data utility, efficiency, identification of malicious and hence controlling the data leakages.
\begin{table}[!htbp]
\centering
    \caption{Comparison for Performance Metrics for various Agents}\label{tabdiffagent}%
        \begin{tabular}{|l|l|l|l|l|l|l|}
			\hline
			No. of & Parameters/ & KNN\cite{afshar2021incorporating} & SVM & NB & QM-MUP\cite{9865138} & \textbf {MAIDS}\\
            Agents & Models & & & & &\\
		    \hline 
            \hline
             \multirow{4}{*}{3000} &PA & 81.83 & 82.50 & 56.66 & 80.54 & 95.33 \\ 
                &PP &  74.95 & 68.00 & 66.47 & 86.45 & 95.24 \\
                &PR &  81.83 & 82.00 & 56.66 & 80.54 & 95.33 \\ 
                &PF &  77.44 & 74.58 & 63.56 & 87.98 & 95.06 \\
            \hline
            \hline
            \multirow{4}{*}{5000} &PA & 86.00 & 82.70 & 58.50 & 90.45 & 93.00 \\ 
                &PP &  79.30 & 68.39 & 69.33 & 92.14 & 93.29 \\
                &PR &  86.00 & 82.69 & 58.50 & 90.56 & 93.00 \\ 
                &PF &  82.16 & 74.86 & 64.95 & 90.42 & 91.61 \\  
			\hline
            \hline
            \multirow{4}{*}{7000} &PA & 86.28 & 85.20 & 59.57 & 94.43 & 93.29 \\ 
                &PP &  80.17 & 72.73 & 65.55 & 91.37 & 93.53 \\
                &PR &  86.28 & 85.28 & 59.57 & 94.54 & 93.28 \\ 
                &PF &  82.67 & 78.51 & 66.18 & 90.16 & 92.01 \\ 
			\hline
            \hline
				\noalign{\smallskip}
	   \end{tabular}
\end{table}

In Fig. \ref{figbaroverall}, a brief comparison of various performance parameters accuracy, precision, recall, and f1 score of proposed model MAIDS is made with state-of-the-art works. Performance parameters are computed for all approaches considering all agents. Observations noted are that firstly, in the case of accuracy, the performance of MAIDS is enhanced in the range 0.87$\%$ to 55.49$\%$ than state-of-the-art works. For precision, improvement lies in the range 6.38$\%$ to 43.15$\%$. In the case of recall a hike of 0.87$\%$ to 55.49$\%$ is noted and for f1-score enhancement lies in the range of 2.68$\%$ to 39.96$\%$. It is evident from this figure that MAIDS performance is dominating over all state-of-the-art works. Therefore, the MAIDS model seems to serve best for the proactive real-time identification of malicious agents and data demands.
\begin{figure*}[!htbp]
\centering
\begin{tikzpicture}[node distance = 1cm,auto,scale=0.95, transform shape]
\begin{axis}[
        height=6.5cm,width=14cm,
        ymin=0, ymax=1,
        ytick={0.4,0.8,1},
        ylabel={Performance Parameters},
        ybar=4*\pgflinewidth,
        bar width=7pt,
        major tick length=0pt,
        ytick style={
            /pgfplots/major tick length=3mm,
        },
        ymajorgrids=true,
        symbolic x coords={PA,PP,PR,PF},
        xtick=data,
        typeset ticklabels with strut,
        enlarge x limits=0.17,
        minor y tick num=5,
        legend columns = -1,
        legend style={
            at={(xticklabel cs:.5)},            
            anchor=north,
            /tikz/every even column/.append style={
                column sep=.3cm,
            },
        },
        set layers,
        cell picture=true,
        extra y ticks=98.23,
        extra y tick labels={},
        extra y tick style={
            ymajorgrids=true,
            ytick style={
                /pgfplots/major tick length=0pt,
            },
            grid style={
                red,
                dashed,
                /pgfplots/on layer=axis foreground,
            },
        },
        mark=none,
        error bars/y dir=both,
        error bars/y explicit,
        error bars/error bar style={
            thick,
        },
    ]
        \addplot [
            fill=lime!20,
        ] coordinates {
            (PA, .9555) 
            (PP, .9575)  
            (PR, .9555)  
            (PF, .9498)
        };
        \addplot [
            fill=orange!20,
        ] coordinates {
            (PA, .9473) 
            (PP, .9000)  
            (PR, .9473)  
            (PF, .9250)
        };
        \addplot [
            fill=cyan!20,
        ] coordinates {
            (PA, .9050) 
            (PP, .8883) 
            (PR, .9050) 
            (PF, .8835)
        };
    
        \addplot [
            fill=pink,
        ] coordinates {
            (PA, .9056) 
            (PP, .8398)  
            (PR, .9056) 
            (PF, .8816) 
        };

        \addplot [
            fill=yellow!60,
        ] coordinates {
            (PA, .8780) 
            (PP, .8321) 
            (PR, .8780)
            (PF, .8432)
        };

         \addplot [
            fill=cyan!10,
        ] coordinates {
            (PA, .8650) 
            (PP, .7482)  
            (PR, .8650) 
            (PF, .8023) 
        };    
        
        \addplot [
            fill=violet!20,
        ] coordinates {
            (PA, .6145) 
            (PP, .6689) 
            (PR, .6145)
            (PF, .6786)
        };        
        \legend{\textbf{MAIDS}, QM-MUP, RF, ANN, KNN, SVM, NB}
    \end{axis}
\end{tikzpicture}
\caption{Comparison for Performance Metrics for all Agents} \label{figbaroverall}
\end{figure*}
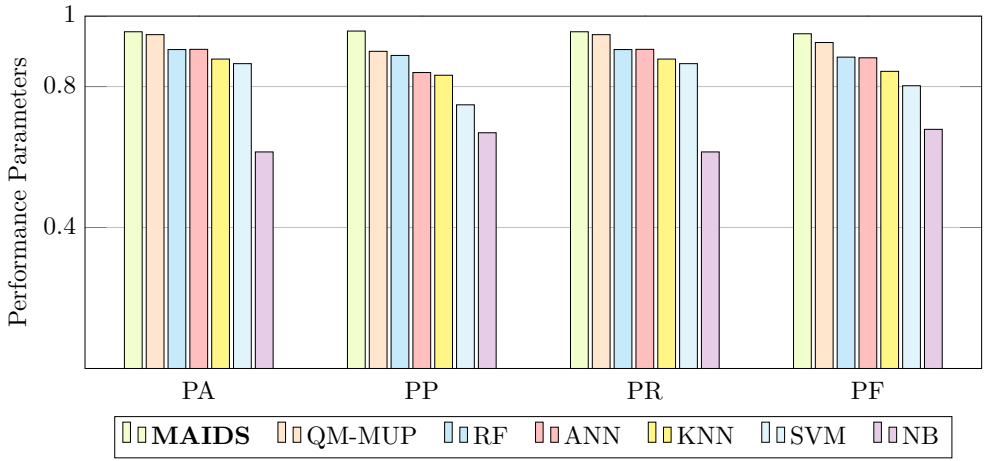
\begin{figure}[!htbp]%
    \centering
\pgfplotstableread{
x y y-max y-min
{SVM-1k} 78 7.0 7.0
{SVM-5k} 84.60 0.60 2.20
{SVM-10k} 85.20 1.30 0.76
{KNN-1k} 82.00 4.85 6.0
{KNN-5k} 85.20 1.7 1.4
{KNN-10k}  86.20 1.6 1.45
{NB-1k}  55.0 5.40 5.0
{NB-5k}  58.60 2.10 3.70
{NB-10k} 58.85 2.6 1.7
{RF-1k} 83.33 5.32 2.83
{RF-5k} 85.80 4.0 0.4
{RF-10k} 89.55 0.95 0.85
{ANN-1k} 82.07 5.59 10.44
{ANN-5k}  86.62 2.86 2.46
{ANN-10k}  87.85 2.71 2.69
{QM-MUP-1k}  88.0 6.73 7.12
{QM-MUP-5k} 88.10 6.63 4.62
{QM-MUP-10k} 88.40 6.33 4.32
{MAIDS-1k} 90.60 4.9 6.6
{MAIDS-5k} 89.90 4.1 3.3
{MAIDS-10k} 93.0 2.6 3.95
}{\differanser}
\begin{tikzpicture}[scale=1.0] 
\begin{axis} [
width  = 0.98*\textwidth,
height = 7cm,
symbolic x coords={{SVM-1k},{SVM-5k},{SVM-10k},{KNN-1k},{KNN-5k},{KNN-10k},{NB-1k},{NB-5k},{NB-10k},{RF-1k},{RF-5k},{RF-10k},{ANN-1k},{ANN-5k},{ANN-10k},{QM-MUP-1k},{QM-MUP-5k},{QM-MUP-10k},{MAIDS-1k},{MAIDS-5k},{MAIDS-10k}},
minor ytick={50,60,70,80,90,100},
yminorgrids,
ylabel={Sucess Rate (\%)},
xtick=data,
ticklabel style = {font=\tiny},
x tick label style={rotate=60,anchor=east},
legend style={at={(0.8,0.28)},anchor=north west,cells={anchor=west},column
sep=1ex}
]
\addplot+[teal, very thick, forget plot,only marks,forget plot] 
plot[very thick, error bars/.cd, y dir=plus, y explicit]
table[x=x,y=y,y error expr=\thisrow{y-max}] {\differanser};
\addplot+[magenta, very thick, only marks,xticklabels=\empty,forget plot] 
plot[very thick, error bars/.cd, y dir=minus, y explicit]
table[x=x,y=y,y error expr=\thisrow{y-min}] {\differanser};
\addplot[only marks,mark=*,mark options={fill=teal,draw=magenta,very thick}] 
table[x=x,y expr=\thisrow{y}] {\differanser};
\addlegendentry{\footnotesize Average}
\addplot[only marks,mark=square*,color=teal!80] 
table[x=x,y expr=\thisrow{y}+\thisrow{y-max}] {\differanser};
\addlegendentry{\footnotesize Max}
\addplot[only marks,mark=square*,color=violet!80] 
table[x=x,y expr=\thisrow{y}-\thisrow{y-min}] {\differanser};
\addlegendentry{\footnotesize Min}
\end{axis} 
\end{tikzpicture}
\caption{Comparative analysis for Success Rate for varying Agents}\label{figlinecahrt}
\end{figure}
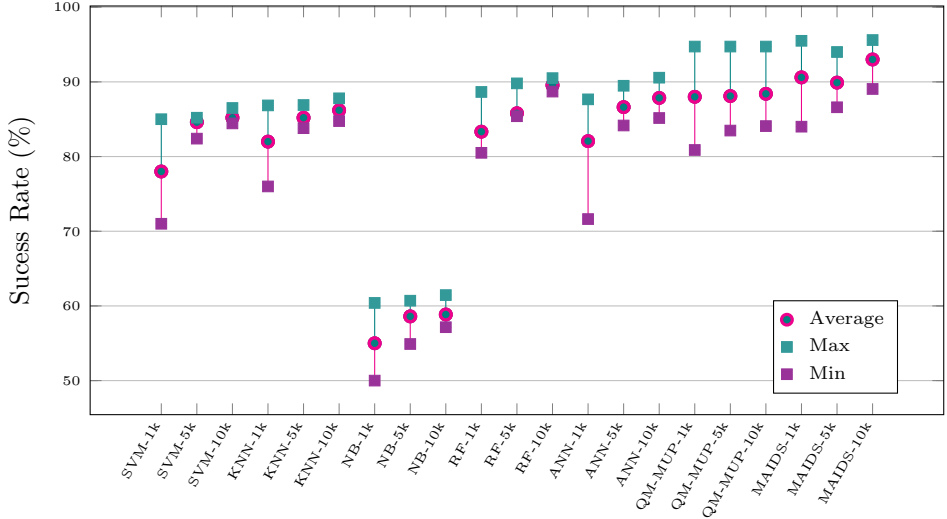

Average success rate (\%) to find out malicious agent, obtained from various state-of-the-art methods such as SVM, KNN, NB, RF, ANN, QM-MUP along with proposed MAIDS model for three different scenarios 1000 (1k) agents, 5000 (5k) agents, and 10000 (10k) agents is presented in the Fig. \ref{figlinecahrt}. It is visible in the line chart that MAIDS is performing best in all three scenarios with different agents that also with a remarkable success rate. The possible reason for this extraordinary performance is the learning optimization capability of MAIDS by employing the computational efficiency of the extreme gradient boosting approach.

Table \ref{taboverall} showcases a comparison of the overall performance parameters and computational complexity of MAIDS with other state-of-the-art models. It is evident from the table that the MAIDS is leading the table with the highest value of performance parameters and lesser complexity. Hence, it can be stated that MAIDS is pioneering malicious agent identification to ensure data security for communication in the cloud platform.
\begin{table*}[!htbp] 
	\centering
		\caption{Parameters and Computational Complexity}
		\label{taboverall}
        \resizebox{0.9\textwidth}{!}{    
			\begin{tabular}{|l|cccc|l|}
				\hline
				Models & Accuracy & Precision & Recall & F1-Score & Computational\\  
			  &  &  &  &  & Complexity\\ 
			  \hline 
			 GAM \cite{5487521} & 59.00 &  55.00 & 96.00 & 70.00 & ${O}(zm + |\sum_{j=1}^{m}d_{j}|)$\\ 
			 DT-ILIS \cite{gupta2019dynamic} & 64.00 & 59.00 & 97.00 & 73.00 & ${O}(z + |\sum_{j=1}^{m}d_{j}|)$   \\ 
			MLPAM \cite{gupta2020mlpam} & 80.00 & 72.00 & 100 & 84.00 & ${O}(|\sum_{j=1}^{m}d_{j}|)$  \\
			{QM-MUP} \cite{9865138} & 94.73 & 90.00 & 94.73 & 92.50 & 
			${O}$($tL{N}$ $N^{\ast})$\\
             \textbf{MAIDS} & 95.55 & 95.30 & 95.50 & 95.20 & 
			${O}$($tmxyN)$
			\\ \hline
				\noalign{\smallskip}
		    \end{tabular}}
\end{table*}
\subsection{Features Comparison}\label{subsec7c}

Table \ref{tabfeature} implies a comparison between numerous framework parameters and computational estimations parameter of MAIDS with state-of-the-art works GAM \cite{5487521}, DT-ILIS \cite{gupta2019dynamic}, MLPAM \cite{gupta2020mlpam}, ABBAC \cite{afshar2021incorporating} and QM-MUP \cite{9865138}. Table showcase that MAIDS is the only model comprising diverse learning rates for model training and retraining purpose which ultimately leads to highly efficient performance. Moreover, MAIDS consists of all other framework parameters as well. Hence it can be stated that MAIDS is discharging the duty of data security for the cloud environment with the highest accuracy to predict malicious agents that too in a proactive manner.
\begin{table*}[!htbp] 
\caption{Features comparison of MAIDS with other state-of-the-art works}
\label{tabfeature}
\centering
\resizebox{1.0\textwidth}{!}{
\begin{tabular}{|l|llllllllll|llll|}
\hline 
\multirow{2}{*}{Scheme} &  \multicolumn{10}{c}{Framework Parameters} &  \multicolumn{4}{|c|}{Estimations}  \\ \cmidrule{2-11} \cmidrule{12-15}
& \textit{$NE$} & \textit{$DP$} & \textit{$A$} & \textit{$DT$} & \textit{$SDS$} & \textit{$SDA$} & \textit{$SDD$} & \textit{$SC$} & \textit{$MEP$} &\textit{$LR$} & $\textit{Acc}$ & $\textit{Pre}$ & $\textit{Rec}$ & $\textit{F1}$ \\   \hline

GAM \cite{5487521} & $\Box$ & $\Box$ & $\boxtimes$ & $\boxtimes$ & $N$ & $N$ & $N$ & $N$ & $\Box$ & $N$ & 59.00 &  55.00 & 96.00 & 70.00  \\ \hline

DT-ILIS \cite{gupta2019dynamic} & $\Box$ & $\Box$ & $\boxtimes$ & $\Box$ & $N$ & $N$ & $N$ & $N$ & $\Box$ & $N$ & 64.00 & 59.00 & 97.00 & 73.00  \\ \hline

MLPAM \cite{gupta2020mlpam} & $\boxtimes$ & $\boxtimes$ & $\boxtimes$ & $\Box$ & $Y$ & $Y$ & $Y$ & $N$ & $\Box$ & $N$ & 84.61 & 82.15 & 84.61 & 81.70  \\ \hline

ABBAC \cite{afshar2021incorporating} & $\Box$ & $\Box$ & $\boxtimes$ & $\boxtimes$ & $N$ & $N$ & $Y$ & $N$ & $\Box$ & $N$ & 87.80 & 83.21 & 87.80 & 84.32  \\ \hline

QM-MUP \cite{9865138} & $\boxtimes$ & $\boxtimes$ & $\boxtimes$ & $\boxtimes$ & $Y$ & $Y$ & $Y$ & $Y$ & $\boxtimes$ & $N$ & 94.73 & 90.00 & 94.73 & 92.50  \\ \hline

MAIDS & $\boxtimes$ & $\boxtimes$ & $\boxtimes$ & $\boxtimes$ & $Y$ & $Y$ & $Y$ & $N$ & $\boxtimes$ & $Y$ & 95.55 & 95.30 & 95.30 & 95.20  \\ \hline
\noalign{\smallskip}
\end{tabular}}
\footnotesize{$\Box$: Single; $\boxtimes$: Multiple; \textit{$NE$}: Non-trusted Entity; \textit{$DP$}: Data Proprietors; \textit{$A$}: Agents; \textit{$SDS$}: Secure Data Storage; \textit{$SDA$}: Secure Data Analysis; \textit{$SDD$}: Secure Data Distribution; \textit{$SC$}: Secure Communication; \textit{$MEP$}: Malicious  Entity Prediction; \textit{$LR$}: Learning Rate; \textit{Acc}: Accuracy; \textit{Pre}: Precision; \textit{Rec}: Recall; \textit{F1}: F1-Score}
\end{table*}
\section{Conclusion}\label{sec8}
    This paper provides a novel comprehensive solution to a crucial data security problem, offering a prescriptive approach for estimating agent behavior by considering multiple security parameters for scrutinizing an agent's behavior.  Thereafter, it implements effective privacy and security practices for the prediction of the malicious agent, proactively using the XG-Boost machine learning approach without compromising data availability and any sort of addition/alteration to the original data, in the cloud environment. The performance evaluation and feature analysis showed that MAIDS ensures high accuracy, precision, recall, F1-score, and success rate and is more protected, efficient, and optimal than existing approaches. The future aim of this work would be to share assembled data in numerous environments and design an adaptive learning governed privacy-preserving mechanism to protect the data for diverse proprietors. 

\section*{Declarations}
\bmhead{Acknowledgments} This work was supported by the National Sun Yat-sen University, Kaohsiung, Taiwan.
\bmhead{Author Contributions} KG, DS, RG, and AKS, contributed to the conception and design of the work, analysis, and interpretation of results, and writing and editing of the manuscript. Implementation, experimentation, data collection, and writing of a first draft of the manuscript were performed by the author, Kishu Gupta.
\bmhead{Funding} Not applicable.
\bmhead{Data and material Availability} The data is available upon reasonable request to the corresponding authors.
\bmhead{Code Availability} The code is available upon reasonable request to the corresponding authors. 
\bmhead{Conflict of interest} The authors have no conflicts of interest to disclose.
\bmhead{Ethical approval} This article does not contain any studies with human participants or animals performed by any of the authors.
\bmhead{Consent to participate} Not applicable. 
\bmhead{Consent for publication} Not applicable.


\bibliographystyle{bst/sn-nature.bst}
\bibliography{reference}
\end{document}